\pdfoutput=1
\documentclass[sigconf]{acmart}
\usepackage{algorithm}
\usepackage[noend]{algpseudocode}
\usepackage{multirow}
\usepackage{booktabs} 
\usepackage{amsmath}
\usepackage{graphicx}
\usepackage[belowskip=-12pt]{caption}
\usepackage{subcaption}
\usepackage{rotating}
\usepackage{adjustbox}
\usepackage{array}
 \usepackage{float}
\usepackage{tabularx}
\usepackage{xcolor}
\usepackage{stfloats}
\usepackage{wrapfig}
\usepackage{listings}
\usepackage{color, colortbl}
\usepackage{balance} 

\usepackage{lscape}
\usepackage{xspace}
\usepackage{framed}
\usepackage{syntax}
\usepackage{parcolumns}
\usepackage{pifont}
\usepackage{soul}

\usepackage[tikz]{bclogo}
\usepackage{enumitem}
\usepackage{soul}
\usepackage{makecell}
\newcommand{\avatar}{\textsc{Avatar}\xspace}
\newcommand{\codenet}{CodeNet\xspace}
\newcommand{\evalplus}{EvalPlus\xspace}
\newcommand{\java}{Java\xspace}
\newcommand{\clang}{C\xspace}

\newcommand{\cpp}{C++\xspace}
\newcommand{\go}{Go\xspace}
\newcommand{\py}{Python\xspace}

\newcolumntype{R}[2]{%
    >{\adjustbox{angle=#1,lap=\width-(#2)}\bgroup}%
    l%
    <{\egroup}%
}

\usepackage[many]{tcolorbox}

\definecolor{problemblue}{RGB}{100,134,158}
\definecolor{idiomsgreen}{RGB}{0,162,0}
\definecolor{exercisebgblue}{rgb}{0,  .69,  .941}
\definecolor{deepgreen}{rgb}{0.0, 0.5, 0.0}
\definecolor{codegreen}{rgb}{0,0.6,0}
\definecolor{codegray}{rgb}{0.5,0.5,0.5}
\definecolor{codepurple}{rgb}{0.58,0,0.82}
\definecolor{backcolour}{rgb}{0.95,0.95,0.92}
\definecolor{redColor}{RGB}{255,0,0}
\definecolor{Gray}{gray}{0.1}

\newtcbtheorem[]{Theorem}{}{
        enhanced,
        sharp corners,
        colback=white,
        colframe=exercisebgblue
    }{thm}

\lstdefinestyle{mystyle}{
	backgroundcolor=\color{backcolour},   
	commentstyle=\color{codegreen},
	keywordstyle=\color{magenta},
	numberstyle=\tiny\color{codegray},
	stringstyle=\color{codepurple},
	basicstyle=\scriptsize,
	breakatwhitespace=false,         
	breaklines=true,                 
	captionpos=b,                    
	keepspaces=false,                 
	numbers=left,                    
	numbersep=5pt,                  
	showspaces=false,                
	showstringspaces=false,
	showtabs=false,                  
	tabsize=2, numbers=left,
    breaklines=true,
    rulecolor=\color{black}
}
\lstdefinelanguage{test}{%
	language     = python,
	breaklines = true,backgroundcolor=\color{white},escapechar=!,rulecolor=\color{black}, breaklines=true,sensitive=true,  numbersep=5pt, xleftmargin=.015\textwidth, frame=tb,label=test
}
\lstdefinelanguage{source}{%
	language     = python,
	breaklines = true,
firstnumber=0,numberfirstline=false,columns=fullflexible,numbers=left,backgroundcolor=\color{white},
    rulecolor=\color{black}, 
    breaklines=true,sensitive=true, numbersep=5pt, xleftmargin=.015\textwidth, label=test
}
\newcommand*\Suppressnumber{%
  \lst@AddToHook{OnNewLine}{%
    \let\thelstnumber\relax%
     \advance\c@lstnumber-\@ne\relax%
    }%
}

\newcommand*\Reactivatenumber{%
  \lst@AddToHook{OnNewLine}{%
   \let\thelstnumber\origthelstnumber%
   \advance\c@lstnumber\@ne\relax}%
}

\NewDocumentCommand{\rangeet}
{ mO{} }{\textcolor{blue}{\textsuperscript{\textit{rangeet}}\textsf{\textbf{\small[#1]}}}}
\NewDocumentCommand{\raju}
{ mO{} }{\textcolor{red}{\textsuperscript{\textit{raju}}\textsf{\textbf{\small[#1]}}}}
\NewDocumentCommand{\reyhan}
{ mO{} }{\textcolor{magenta}{\textsuperscript{\textit{Reyhan}}\textsf{\textbf{\small[#1]}}}}
\NewDocumentCommand{\ali}
{ mO{} }{\textcolor{purple}{\textsuperscript{\textit{Ali}}\textsf{\textbf{\small[#1]}}}}
\NewDocumentCommand{\rahul}
{ mO{} }{\textcolor{red}{{\textit{Rahul:~#1}}}}

\definecolor{diffstart}{named}{codegreen}
\definecolor{diffincl}{named}{redColor}

\lstdefinestyle{customc}{
	belowcaptionskip=1\baselineskip,
	breaklines=false,
	frame= single,
	breaklines = true,
	xleftmargin=\parindent,
	language= Python,
	showstringspaces=false,
	basicstyle=\footnotesize\ttfamily,
	keywordstyle=\bfseries\color{green!40!black},
	commentstyle=\itshape\color{purple!40!black},
	identifierstyle=\color{blue},
	stringstyle=\color{codegreen},
	backgroundcolor=\color{gray!4}
}

\DeclareCaptionFormat{listing}{#3}
\newcommand{\change}[1]{#1}

\graphicspath{{./figures/}}

\newcommand{\cardinal}[1]{\textit{Cardinal}}

\newcommand*\circled[1]{\tikz[baseline=(char.base)]{
		\node[shape=circle,fill,inner sep=.2pt] (char) {\textcolor{white}{#1}};}}

\newcounter{rqs}
\stepcounter{rqs}
\newcommand{\sect}[1]{\S\ref{sec:#1}\xspace}

\newcounter{NumObservations}
\stepcounter{NumObservations}
\definecolor{shadecolor}{rgb}{.9,.9,.9}

\definecolor{msftBlue}{RGB}{0,164,239}
\definecolor{msftGreen}{RGB}{127,186,0}
\definecolor{msftYello}{RGB}{255,185,0}
\usepackage{bigstrut}
\newenvironment{findingBox}[2]{%
	\begin{tcolorbox}[colframe=black!90,colback=exercisebgblue!10,boxrule=.5pt, left=1pt, right = 1pt, top=0pt, bottom=0pt, size=small]{\textbf{Finding #1:} #2} 
}{%
	\end{tcolorbox}
}

\newcommand{\ignore}[1]{{}}

\newcommand\bi{\begin{itemize}[leftmargin=*, itemindent=0pt, wide=0pt]}
\newcommand\ei{\end{itemize}}
\newcommand\be{\begin{enumerate}[leftmargin=*, itemindent=0pt, wide=0pt]}
\newcommand\ee{\end{enumerate}}

\definecolor{vlcolor}{rgb}{0.9,0.1,0.1}

\newcommand{\squarebox}[1]{
  \tikz[baseline=(char.base)]{
    \node[shape=rectangle,draw,inner sep=2pt, blue] (char) {#1};
  }
}

\AtBeginDocument{%
  }

\copyrightyear{2024}
\acmYear{2024}
\setcopyright{rightsretained}
\acmConference[ICSE '24]{2024 IEEE/ACM 46th International Conference on Software Engineering}{April 14--20, 2024}{Lisbon, Portugal}
\acmBooktitle{2024 IEEE/ACM 46th International Conference on Software Engineering (ICSE '24), April 14--20, 2024, Lisbon, Portugal}\acmDOI{10.1145/3597503.3639226}
\acmISBN{979-8-4007-0217-4/24/04}

\acmPrice{15.00}

  {\list{}{\leftmargin=#1\rightmargin=#1\footnotesize}\item[]}%
  {\endlist}

\begin{document}


\title{Lost in Translation: A Study of Bugs Introduced by Large Language Models while Translating Code}

\author{Rangeet Pan}
\authornote{Both authors contributed equally to this research.}
\email{rangeet.pan@ibm.com}
\affiliation{
  \institution{IBM Research}
  \city{Yorktown Heights}
  \state{NY}
  \country{USA}
  \postcode{10606}
}
\author{Ali Reza Ibrahimzada}
\authornote{Author was an intern at IBM Research at the time of this work.}
\authornotemark[1]
\email{alirezai@illinois.edu}
\affiliation{
  \institution{\mbox{University~of~Illinois~Urbana-Champaign}}
  \city{Champaign}
  \state{IL}
  \country{USA}
}
\author{Rahul Krishna}
\email{rkrsn@ibm.com}
\affiliation{
  \institution{IBM Research}
  \city{Yorktown Heights}
  \state{NY}
  \country{USA}
  \postcode{10606}
}
\author{Divya Sankar}
\email{divya.sankar@ibm.com}
\affiliation{
  \institution{IBM Research}
  \city{Yorktown Heights}
  \state{NY}
  \country{USA}
  \postcode{10606}
}
\author{Lambert Pougeum Wassi}
\email{lambert.pouguem.wassi@ibm.com}
\affiliation{
  \institution{IBM Research}
  \city{Yorktown Heights}
  \state{NY}
  \country{USA}
  \postcode{10606}
}
\author{Michele Merler}
\email{mimerler@us.ibm.com}
\affiliation{
  \institution{IBM Research}
  \city{Yorktown Heights}
  \state{NY}
  \country{USA}
  \postcode{10606}
}
\author{Boris Sobolev}
\email{bsobolev@ibm.com}  
\affiliation{
  \institution{IBM Research}
  \city{Yorktown Heights}
  \state{NY}
  \country{USA}
  \postcode{10606}
}
\author{Raju Pavuluri}
\email{pavuluri@us.ibm.com}  
\affiliation{
  \institution{IBM Research}
  \city{Yorktown Heights}
  \state{NY}
  \country{USA}
  \postcode{10606}
}
\author{Saurabh Sinha}
\email{sinhas@us.ibm.com}  
\affiliation{
  \institution{IBM Research}
  \city{Yorktown Heights}
  \state{NY}
  \country{USA}
  \postcode{10606}
}
\author{Reyhaneh Jabbarvand}
\email{reyhaneh@illinois.edu}  
\affiliation{
  \institution{\mbox{University~of~Illinois~Urbana-Champaign}}
  \city{Champaign}
  \state{IL}
  \country{USA}
}

\renewcommand{\shortauthors}{Pan et al.}
\begin{abstract}
Code translation aims to convert source code from one programming language (PL) to another. Given the promising abilities of large language models (LLMs) in code synthesis, researchers are exploring their potential to automate code translation. The prerequisite for advancing the state of LLM-based code translation is to understand their promises and limitations over existing techniques. To that end, we present a large-scale empirical study to investigate the ability of general LLMs and code LLMs for code translation across pairs of different languages, including \clang, \cpp, \go, \java, and \py. Our study, which involves the translation of 1,700 code samples from three benchmarks and two real-world projects, reveals that LLMs are yet to be reliably used to automate code translation---with correct translations ranging from 2.1\% to 47.3\% for the studied LLMs. Further manual investigation of \textit{unsuccessful} translations identifies 15 categories of translation bugs. We also compare LLM-based code translation with traditional non-LLM-based approaches. Our analysis shows that these two classes of techniques have their own strengths and weaknesses. Finally, insights from our study suggest that providing more context to LLMs during translation can help them produce better results. To that end, we propose a prompt-crafting approach based on the symptoms of erroneous translations; this improves the performance of LLM-based code translation by 5.5\% on average. Our study is the first of its kind, in terms of scale and breadth, that provides insights into the current limitations of LLMs in code translation and opportunities for improving them. Our dataset---consisting of 1,700 code samples in five PLs with 10K+ tests, 43K+ translated code, 1,748 manually labeled bugs, and 1,365 bug-fix pairs---can help drive research in this area.
\end{abstract}



\maketitle

\vspace{-5pt}
\section{Introduction}
\label{sec:intro}

Code translation, source-to-source compilation, or transpilation, entails transforming a piece of code from one programming language (PL) to another, while preserving the original functionality. Code translation has many use cases, such as modernizing enterprise applications~\cite{krishna2021transforming,perez2021software,settu2013cloud,echeverria2015legacy,kalia2021mono2micro}, migrating legacy software in proprietary PLs to cloud-native applications implemented in general-purpose PLs~\cite{haugeland2021migrating,kazanavivcius2019migrating,pei2020towards,zhang2009migrating,gholami2017challenges,bergmayr2013migrating,nitin2022cargo}, and facilitating the training of models for better code synthesis~\cite{chen2023exploring,heo2022end,subedi2021application,silva2023mufin}. Translating the software/code to a modern 
PL can significantly reduce maintenance effort, improve overall reliability, and boost non-functional properties such as security and performance~\cite{thones2015microservices,intro1,intro2,intro3,intro4,intro5}. 

Due to the importance and benefits of code translation, several techniques have been developed to automate reliable translation between different PLs~\cite{roziere2020unsupervised,roziere2021leveraging,liu2023syntax,szafraniec2022code,wen2022babeltower,ting2023codestylist,hong2023improving,weisz2021perfection,weisz2022better,nguyen2015divide,chen2018tree,c2go,c2rust,java2csharp},
including those leveraging \textit{large language models} (LLMs) for code translation~\cite{jana2023attention,gong2023adelt,roziere2020unsupervised,roziere2021leveraging,weisz2021perfection,sun2023transcoder}. 
Although prior research has shown the potential of using LLMs for code translation, there is a dearth of research on understanding and cataloging their limitations for this task.  
This is an important undertaking because code translation is a complex task that requires LLMs to understand code syntax (to generate syntactically correct code) and semantics (to preserve functionality during translation) simultaneously. However, research has shown that without providing adequate context to LLMs via prompt crafting, they may only serve as ``next code token'' predictors, without understanding the overall task~\cite{white2023prompt,abukhalaf2023codex,jiang2023self,zhuo2023robustness}. 

In this work, we perform a large-scale empirical study to understand \change{the promises and limitations of LLM-based code translation, and compare them with existing non-LLM-based translation approaches.} 
We also perform a preliminary investigation of how 
\change{providing more context about incorrect translations 
improves the results}. Our study answers the following research questions:

\begin{description}[leftmargin=0pt]
\item[RQ1:] \hspace{-3pt} \textbf{Effectiveness in Code Translation (\sect{vanilla-prompting}).} 
(\textit{RQ1.1}) How do state-of-the-art general and code LLMs perform 
in code translation? 
(\textit{RQ1.2}) 
What are the outcomes of unsuccessful translations?

\item[RQ2:] \textbf{LLM-Based Translation Bugs (\sect{taxonomy}).} 
(\textit{RQ2.1}) What are the different types of underlying root causes (translation bugs) for unsuccessful translations?
(\textit{RQ2.2}) How prevalent are these bugs in unsuccessful translations? 
(\textit{RQ2.3}) How do translation bugs in \textit{real-world projects} differ from those in \textit{crafted benchmarks}? 
\change{\item[RQ3:] \textbf{Comparison with Alternative Approaches (\sect{comparison}).} How do state-of-the-art non-LLM-based techniques perform in code translation and what types of translation bugs do they introduce?} 

\item[RQ4:] \textbf{Mitigating Translation Bugs (\sect{prompt-engineering}}). To what extent do the proposed \textit{prompt-crafting techniques} resolve 
translation bugs?

\end{description}

\noindent To investigate the RQs, we collected 1,700 executable code samples from \textit{three} well-known datasets (\codenet~\cite{puri2codenet}, \avatar~\cite{ahmad2021avatar}, and \evalplus~\cite{liu2023your}) and \textit{two} open-source projects (Apache Commons CLI~\cite{apachecommonscli} and \py Click~\cite{click}), covering \textit{five} PLs (\clang, \cpp, \go, \java, and \py). To perform translation, we selected \textit{seven} LLMs: GPT-4~\cite{openai2023gpt4}, \textit{three} open-source LLMs from the Hugging Face Open LLM Leaderboard~\cite{hfleaderboard} (Llama~2~\cite{llama2}, TheBloke-Vicuna~\cite{wizard}, and TheBloke-Airoboros~\cite{airoboros}), and \textit{three} recent code LLMs 
(StarCoder~\cite{li2023starcoder}, CodeGeeX~\cite{zheng2023codegeex}, and CodeGen~\cite{nijkamp2022codegen}). 

We performed 43,379 translations across all LLMs, measuring translation success against the tests provided with the code samples. This produced 11.94\% successful translations on average (median 5.3\%), with GPT-4 (47.3\% success rate) and StarCoder (14.5\% success rate) being the best-performing models (details in \S\ref{subsec:llmeffectiveness}). 
On real-world projects, the LLMs were largely ineffective, with success rates of 8.1\% for GPT-4 and 0\% for the rest of the models.

We also conducted a systematic study to understand the root causes of unsuccessful translations and create a taxonomy of translation bugs. \change{The process involved eight human labelers and took ~630 person-hours in total, focusing on 1,725 buggy translations by GPT-4. It was conducted in two phases: in Phase~1, a draft taxonomy was created from buggy translations for one language pair; in Phase~2, the taxonomy was used for other language pairs and extended, if needed, to label buggy translations (details in \S\ref{sec:taxonomy}).}
The resulting bug taxonomy is structured into \change{$15$} categories and \change{five} groups (details in \S\ref{subsec:taxonomy}).
Some notable findings are: (1) identifying suitable data type in the target PL that preserves the source behavior is challenging, (2) identifying equivalent APIs in the target language or implementing the API functionality can introduce bugs, and (3) replacing language-specific features, such as method overloading and annotations, can be challenging, especially in real-world projects.
Another important dimension of this study is comparing LLM-based translation with existing non-LLM-based techniques, namely, CxGo~\cite{c2go}, C2Rust~\cite{c2rust}, and JavaToCsharp~\cite{java2csharp}.\footnote{We also considered other approaches (mppSMT~\cite{nguyen2015divide}, Tree2Tree~\cite{chen2018tree}, and Sharpen~\cite{sharpen}) but could not compare against them due to lack of tool/artifact availability (\S\ref{sec:comparison}).} The comparison shows that LLM- and non-LLM-based translation techniques provide different and unique advantages, suggesting that an ultimate solution for code translation should combine both techniques (details in \S\ref{sec:comparison}).

\change{Our study reveals that, often, providing only the source code may be insufficient for achieving correct code translation. \change{To that end} (and also motivated
by recent research on LLM-based bug repair~\cite{joshi2023repair, xia2022less}), we propose an iterative prompting approach
that incorporates} additional informative context in prompts corresponding to the previously unsuccessful translation, including the code,  
stack trace, error message from failing execution, and/or test input and expected output from failing test cases.
Our results show prompt crafting increases the success rate by 5.5\%, on average, across the studied LLMs, with the largest improvement, of 12\%, occurring for GPT-4 (details in \S\ref{sec:prompt-engineering}).
\change{Although these results are encouraging, they indicate considerable scope for improvement, likely through a combination of program analysis techniques and LLMs.}

To our knowledge, we are the first to (1)
provide a systematic bug taxonomy facilitating a deeper understanding of error modalities in LLM-based code translation, (2) 
study translation of real-world projects, (3) 
investigate the effectiveness of prompt crafting in 
mitigating translation bugs, \change{and (4) compare non-LLM and LLM-based translation approaches}. Our key contributions are:

\begin{itemize}[noitemsep, nolistsep, leftmargin=*]
    \item \textbf{A comprehensive evaluation of LLM-based code translation.} We perform a large-scale evaluation of code translation using multiple general and code LLMs. 
    We consider the recently released LLMs, and our evaluation includes real-world projects in addition to three crafted benchmarks. 
    
    \item \textbf{A taxonomy of translation bugs.} Our study offers the first taxonomy of bugs introduced by LLMs during code translation. 
    We also compare the nature of these bugs in LLM and non-LLM-based approaches to understand their strengths and weaknesses.
     \item \textbf{Prompt crafting to enhance code translation.} A set of heuristics for prompt crafting that provides proper contexts to LLMs to improve their effectiveness in code translation. 
    \item \textbf{Artifacts.} Our artifacts, including manual labeling and automation scripts for evaluating LLMs, are publicly available~\cite{website}.
    
\end{itemize}

%
\vspace{-1pt}
\section{Empirical Setup}
\label{sec:empirical-setup}

\noindent \textbf{Subject LLM Selection.}
General LLMs
are pre-trained on textual data, including natural language and code, and can be used for a variety of tasks. 
In contrast, code LLMs are specifically pre-trained to automate code-related tasks. Due to the empirical nature of this work, we were interested in assessing the effectiveness of both LLM categories in code translation. For code LLMs, we selected the top three models released recently (in $2023$), namely CodeGen~\cite{nijkamp2022codegen}, StarCoder~\cite{li2023starcoder}, and CodeGeeX~\cite{zheng2023codegeex}. For general LLMs, we selected the top three models with size $20$B parameters or less from the Hugging Face Open LLM Leaderboard~\cite{hfleaderboard}.\footnote{The Open LLM Leaderboard ranking is quite dynamic, and our selection is drawn from the ranking at the time of our experimentation.} The constraint on the number of parameters was imposed by our computing resources, resulting in the selection of Llama 2~\cite{llama2}, TheBloke-Airoboros~\cite{airoboros}, and TheBloke-Vicuna~\cite{wizard}. We also included GPT-4~\cite{openai2023gpt4} in our study. 
Table~\ref{table:models} summarizes characteristics of the selected LLMs.

\textbf{Subject PLs Selection.} 
We used the following criteria to select the subject PLs: (1) popularity of the language based on the TIOBE index~\cite{tiobe}, (2) inclusion of different programming paradigms, e.g., procedural, object-oriented, and functional, and (3) availability of high-quality datasets in the given PL. To make the manual effort involved in taxonomy construction manageable, we selected five PLs that met the inclusion criteria---\clang, \cpp, \go, \java, and \py.

\textbf{Dataset Collection and Pre-Processing.}
To ensure the comprehensiveness of our findings and claims in understanding the nature of LLM translation bugs, we were interested in datasets used in prior studies as well as real-world projects. The former consists of small programs, likely to be less challenging for LLMs to translate, and the latter assesses the complexity of LLM translation in real-world settings. 









\begin{table}[t]
  \centering
  \footnotesize
  \setlength\tabcolsep{0.55pt}
  \caption{Overview of subject LLMs. TB: TheBloke.}
\vspace*{-10pt}
\resizebox{\columnwidth}{!}{
    \begin{tabular}{|l|c|c|c|c|c|c|c|}
    \hline
    \multicolumn{1}{|c|}{\textbf{Modality}} & \multicolumn{3}{c|}{\textbf{Code}} & \multicolumn{4}{c|}{\textbf{Text}} \bigstrut\\
    \hline
    \multicolumn{1}{|c|}{\textbf{Models}}  & CodeGen  &CodeGeeX & StarCoder  & GPT-4 & Llama 2  & TB-Airboros & TB-Vicuna  \bigstrut\\
    \hline
    \hline
    \textbf{Size}  & 16B   & 13B   & 15.5B & -     & 13B   & 13B   & 13B \bigstrut\\
    \hline
    \textbf{Context Window}  & 2048   & 2048   & 2048 & 8192     & 4096   & 2048   & 2048 \bigstrut\\
    \hline
    \textbf{Release Date} & May'23 & Mar'23 & May'23 & Mar'23 & Jul'23 & May'23 & May'23 \bigstrut\\
    \hline
    \end{tabular}}%
 \label{table:models}
\vspace*{-2em}
\end{table}%

The first six columns of Table~\ref{table:dataset} present the selected datasets and statistics about them (more information in the artifact website~\cite{website}). These datasets are accompanied by test cases to validate code translation. For \codenet and \avatar, the tests are input data and expected output, while \evalplus and real-world projects have unit tests (JUnit and pytest). For \evalplus, we manually translated and verified the corresponding pytests to JUnit tests. The translation of real-world projects never reached test execution, as it produced syntactically incorrect code (more discussion in \S\ref{sec:vanilla-prompting}). 
\begin{table*}[t]
  \centering
  \caption{\change{Performance of subject LLMs in translating code from different studied datasets. The best performance by general and code LLMs are highlighted in teal and violet, respectively. The final performance is computed over the average of each dataset.}}
  \vspace*{-11pt}
  \label{table:dataset}
  \setlength\tabcolsep{1.5pt}
    \resizebox{\textwidth}{!}{
    \begin{tabular}{l|c|c|c|c|c|c|c|c|c|c|c|c}
    \hline
    \multirow{2}{*}{\textbf{Dataset}} &
    \multirow{3}{*}{\textbf{Language}} &
    \multirow{3}{*}{\textbf{Samples}} &
    \multirow{2}{*}{\textbf{\#Tests}}& 
    \multirow{3}{*}{\textbf{Language}} &
    \multirow{2}{*}{\textbf{\#Translations}}&
    \multicolumn{7}{c}{\textbf{\change{\% Successful Translations}}} 
    \\\cline{7-13}
    \multicolumn{1}{l|}{} &
    \multirow{-3}{*}{\textbf{Source}} &
    \multirow{-3}{*}{\textbf{Source}} &
    \multicolumn{1}{c|}{} &
    \multirow{-3}{*}{\textbf{Target}} &
    &
    \multicolumn{1}{c|}{{\textbf{CodeGen}}} &
    \multicolumn{1}{c|}{{\textbf{CodeGeeX}}} &
    \multicolumn{1}{c|}{{\textbf{StarCoder}}} &
    \multicolumn{1}{c|}{{\textbf{GPT-4}}} &
    \multicolumn{1}{c|}{{\textbf{Llama 2}}} &
    \multicolumn{1}{c|}{{\textbf{TB-Airoboros}}} &
    \multicolumn{1}{c}{{\textbf{TB-Vicuna}}}
    \\
    \hline
    \hline
    \multirow{6}{*}{\codenet~\cite{puri2codenet}} & \multirow{1}{*}{C} & \multirow{1}{*}{200} & \multirow{1}{*}{200} &
    C++, Go, Java, Python &
    800&23.4\%&14.9\%&
    {\color{violet}\textbf{42.0\%}}
    &
    {\color{teal}\textbf{83.0\%}}
    &14.9\%&18.8\%&4.4\%
    \\
    \cline{2-13}
    & \multirow{1}{*}{C++} & \multirow{1}{*}{200} & \multirow{1}{*}{200} & 
    C, Go, Java, Python &
    800&14.0\%&3.6\%&{\color{violet}\textbf{39.1\%}}
    &
    {\color{teal}\textbf{80.0\%}}
    &9.5\%&8.3\%&3.4\%
    \\
    \cline{2-13}
    & \multirow{1}{*}{Go} & \multirow{1}{*}{200} & \multirow{1}{*}{200} & 
    C, C++, Java, Python &
    800&14.3\%&5.9\%&
    {\color{violet}\textbf{42.0\%}}
    &
    {\color{teal}\textbf{85.5\%}}
    &16.9\%&6.6\%&0.9\%
    \\
    \cline{2-13}
    & \multirow{1}{*}{Java} & \multirow{1}{*}{200} & \multirow{1}{*}{200} & 
    C, C++, Go, Python &
    800&21.3\%&10.3\%&
    {\color{violet}\textbf{30.3\%}}
    &
    {\color{teal}\textbf{81.3\%}}
    &13.9\%&6.5\%&0.1\%
    \\
    \cline{2-13}
    & \multirow{1}{*}{Python} & \multirow{1}{*}{200} & \multirow{1}{*}{200} & 
    C, C++, Go, Java &
    800&17.5\%&7.3\%&
    {\color{violet}\textbf{33.3\%}}&
    {\color{teal}\textbf{79.9\%}}
    &11.0\%&6.5\%&1.0\%
    \\
    \hline
    \cellcolor[rgb]{0.851 0.851 0.851}Total/Average (\codenet)
    &\cellcolor[rgb]{0.851 0.851 0.851} - 
    & \cellcolor[rgb]{0.851 0.851 0.851}1,000 
    & \cellcolor[rgb]{0.851 0.851 0.851}1,000 
    & \cellcolor[rgb]{0.851 0.851 0.851}-  
    & \cellcolor[rgb]{0.851 0.851 0.851}4,000
    &\cellcolor[rgb]{0.851 0.851 0.851}18.1\%
    &\cellcolor[rgb]{0.851 0.851 0.851}8.4\%
    &\cellcolor[rgb]{0.851 0.851 0.851}{\color{violet}\textbf{37.3\%}}
    &\cellcolor[rgb]{0.851 0.851 0.851}{\color{teal}\textbf{82.0\%}}
    &\cellcolor[rgb]{0.851 0.851 0.851}13.2\%
    &\cellcolor[rgb]{0.851 0.851 0.851}9.3\%
    &\cellcolor[rgb]{0.851 0.851 0.851}2.0\%
    \\
    \hline
    \multirow{2}{*}{\avatar~\cite{ahmad2021avatar}} & \multirow{1}{*}{Java} & \multirow{1}{*}{249} &\multirow{2}{*}{6,255} &  C, C++, Go, Python 
    & $996$&8.1\%&1.8\%&
    {\color{violet}\textbf{11.9\%}}&
    {\color{teal}\textbf{70.8\%}}
    &1.8\%&5.0\%&0.0\%
    \\
    \cline{2-3}\cline{5-13}        & \multirow{1}{*}{Python} & \multirow{1}{*}{250} &  
    & C, C++, Go, Java   
    & $1,000$
    &3.8\%&1.6\%&
    {\color{violet}\textbf{14.2\%}}&
    {\color{teal}\textbf{52.2\%}}
    &4.7\%&0.9\%&0.9\%
    \\
    \hline
    \cellcolor[rgb]{0.851 0.851 0.851}Total/Average (\avatar)
    & \cellcolor[rgb]{0.851 0.851 0.851}- 
    &\cellcolor[rgb]{0.851 0.851 0.851} 499 
    &\cellcolor[rgb]{0.851 0.851 0.851} 6,255 
    &\cellcolor[rgb]{0.851 0.851 0.851} -  
    & \cellcolor[rgb]{0.851 0.851 0.851}1,996
    &\cellcolor[rgb]{0.851 0.851 0.851}5.9\%
    &\cellcolor[rgb]{0.851 0.851 0.851}1.7\%
    &\cellcolor[rgb]{0.851 0.851 0.851}{\color{violet}\textbf{13.0\%}}
    &\cellcolor[rgb]{0.851 0.851 0.851}{\color{teal}\textbf{61.5\%}}
    &\cellcolor[rgb]{0.851 0.851 0.851}3.2\%
    &\cellcolor[rgb]{0.851 0.851 0.851}3.0\%
    &\cellcolor[rgb]{0.851 0.851 0.851}0.4\%
    \\
    \hline
    \multirow{1}{*}{\evalplus~\cite{liu2023your}} & \multirow{1}{*}{Python} & \multirow{1}{*}{164} & \multirow{1}{*}{2,682} & Java  
    & $164$
    &16.5\%&3.7\%&{\color{violet}\textbf{22.0\%}}&{\color{teal}\textbf{79.3\%}}&1.2\%&14.0\%&7.9\%
    \\
    \hline
    \multirow{1}{*}{Commons CLI~\cite{apachecommonscli}} & \multirow{1}{*}{Java} & \multirow{1}{*}{22} & \multirow{1}{*}{310} & Python  
    & $22$
    &0.0\%&0.0\%&0.0\%&
    {\color{teal}\textbf{13.6\%}}&0.0\%&0.0\%&0.0\%
    \\
    \hline
    \multirow{1}{*}{Click~\cite{click}} & \multirow{1}{*}{Python} & \multirow{1}{*}{15} & \multirow{1}{*}{611} & Java  
    & $15$
&0.0\%&0.0\%&0.0\%&0.0\%&0.0\%&0.0\%&0.0\%
    \\
    \hline
    \hline
    \multicolumn{1}{l|}{\cellcolor[rgb]{0.851 0.851 0.851}Total/Average (All)} 
    &\cellcolor[rgb]{0.851 0.851 0.851}- 
    &\cellcolor[rgb]{0.851 0.851 0.851}1,700 
    &\cellcolor[rgb]{0.851 0.851 0.851}10,858 
    &\cellcolor[rgb]{0.851 0.851 0.851}- 
    &\cellcolor[rgb]{0.851 0.851 0.851}6,197
    &\cellcolor[rgb]{0.851 0.851 0.851}8.1\%
    &\cellcolor[rgb]{0.851 0.851 0.851}2.8\%
    &\cellcolor[rgb]{0.851 0.851 0.851}{\color{violet}\textbf{14.5\%}}
    &\cellcolor[rgb]{0.851 0.851 0.851}{\color{teal}\textbf{47.3\%}}
    &\cellcolor[rgb]{0.851 0.851 0.851}3.5\%
    &\cellcolor[rgb]{0.851 0.851 0.851}5.3\%
    &\cellcolor[rgb]{0.851 0.851 0.851}2.1\%
    \\
    \hline
    \end{tabular}
    }\\[1em]%
\vspace*{-18pt}
\end{table*}%

For real-world projects, we focused on \java and \py, the most popular languages among our subject PLs. Our goal was to translate reasonably complex and well-maintained software exclusively written in \java or \py.
To that end, we selected projects available in both PLs providing APIs for command-line processing and selected Apache Commons CLI~\cite{apachecommonscli} (\java) and Click~\cite{click} (\py). 
To
fit the source language code into the limited LLM context window,
we broke them down into classes and files and removed all comments. 


\textbf{Compute Resources.} To perform inference on all subject LLMs, we used 16 A100 80GB memory GPUs. For evaluating the generated translations, we used Python 3.10, g++ 11, GCC Clang 14.0, Java 11, Go 1.20, Rust 1.73, and .Net 7.0.14 for Python, C++, C, Java, Go, Rust, and C\#, respectively.

%
\vspace{-5pt}
\section{LLM-Based Code Translation}
\label{sec:vanilla-prompting}
\begin{table*}
  \centering
  \footnotesize
  \setlength\tabcolsep{1.5pt}
  \vspace{14pt}
  \caption{Breakdown of the unsuccessful translations produced by subject LLMs based on outcome. All values are in \%.}
  \vspace{-10pt}
  \label{table:breakdown}
    \begin{tabular}{|l|r|r|r|r|r|r|r|r|r|r|r|r|r|r|r|r|r|r|r|r|r|}
\cline{1-21}    \textbf{Source Language} & \multicolumn{4}{c|}{\textbf{C}} & \multicolumn{4}{c|}{\textbf{C++}} & \multicolumn{4}{c|}{\textbf{Go}} & \multicolumn{4}{c|}{\textbf{Java}} & \multicolumn{4}{c|}{\textbf{Python}} & \multicolumn{1}{c}{\multirow{2}[3]{*}{\cellcolor[rgb]{ .502,  .502,  .502}\textcolor[rgb]{ 1,  1,  1}{\textbf{}}}} \bigstrut\\
\cline{1-21}    \textbf{Target language} & \multicolumn{1}{c|}{\textbf{C++}} & \multicolumn{1}{c|}{\textbf{Go}} & \multicolumn{1}{c|}{\textbf{Java}} & \multicolumn{1}{c|}{\textbf{Python}} & \multicolumn{1}{c|}{\textbf{C}} & \multicolumn{1}{c|}{\textbf{Go}} & \multicolumn{1}{c|}{\textbf{Java}} & \multicolumn{1}{c|}{\textbf{Python}} & \multicolumn{1}{c|}{\textbf{C}} & \multicolumn{1}{c|}{\textbf{C++}} & \multicolumn{1}{c|}{\textbf{Java}} & \multicolumn{1}{c|}{\textbf{Python}} & \multicolumn{1}{c|}{\textbf{C}} & \multicolumn{1}{c|}{\textbf{C++}} & \multicolumn{1}{c|}{\textbf{Go}} & \multicolumn{1}{c|}{\textbf{Python}} & \multicolumn{1}{c|}{\textbf{C}} & \multicolumn{1}{c|}{\textbf{C++}} & \multicolumn{1}{c|}{\textbf{Go}} & \multicolumn{1}{c|}{\textbf{Java}} & \multirow{-2}{*}{\cellcolor[rgb]{ .502,  .502,  .502}\textcolor[rgb]{ 1,  1,  1}{\textbf{Total}}} \bigstrut\\
\hline
\hline
    Compilation Errors & 68.9 & 93.5 & 76.4 & 56.9 & 93.2 & 94.6 & 77.0 & 61.6 & 86.7 & 83.3 & 82.4 & 55.9 & 82.4 & 78.4 & 96.6 & 57.4 & 79.9 & 73.4 & 86.0 & 72.4 & \cellcolor[rgb]{ 0.851,  0.851,  0.851}\textbf{77.8} \bigstrut\\
    \hline
    Runtime Errors & 9.4 & 2.3 & 10.7 & 21.9 & 0.1 & 1.2 & 11.2 & 22.9 & 0.2 & 0.2 & 12.7 & 19.3 & 1.2 & 0.4 & 0.8 & 27.1 & 0.4 & 0.4 & 10.0 & 14.8 & \cellcolor[rgb]{ 0.851,  0.851,  0.851}\textbf{8.4} \bigstrut\\
    \hline
    Functional Errors & 20.5 & 3.7 & 13.0 & 20.6 & 6.7 & 4.1 & 11.7 & 15.1 & 12.9 & 16.3 & 4.7 & 24.6 & 15.8 & 19.9 & 2.5 & 15.1 & 19.0 & 24.8 & 3.9 & 12.5 & \cellcolor[rgb]{ 0.851,  0.851,  0.851}\textbf{13.4} \bigstrut\\
    \hline
    Non-terminating Execution & 1.3 & 0.4 & 0.0 & 0.5 & 0.0 & 0.2 & 0.1 & 0.4 & 0.2 & 0.2 & 0.2 & 0.2 & 0.7 & 1.3 & 0.1 & 0.3 & 0.8 & 1.3 & 0.1 & 0.3 & \cellcolor[rgb]{ 0.851,  0.851,  0.851}\textbf{0.4} \bigstrut\\
\hline    
\end{tabular}%
\vspace*{-10pt}
\end{table*}%

We prompted each subject LLMs with 6,197 translation problems corresponding to 31 translation pairs shown in Table~\ref{table:dataset}, i.e., 20~pairs from \codenet, eight pairs from \avatar, and one pair each for EvalPlus, Commons CLI, and Click.
Through RQ1, we evaluate the effectiveness of LLMs in code translation (RQ1.1) and 
the outcomes of incorrect translations
(RQ1.2).  

\subsection{Effectiveness of LLMs in Code Translation}
\label{subsec:llmeffectiveness}



We refer to the LLM prompting in this experiment as \textit{vanilla prompting}, where each prompt contains four pieces of information: (1) instructions in natural language to perform the translation task,
(2) source language (\texttt{\small\textcolor[RGB]{134,60,132}{\$SOURCE_LANG}}), (3) target language (\texttt{\small\textcolor{red!70}{\$TARGET_LANG}}), and (4) the code to be translated (\texttt{\small\textcolor[RGB]{65, 124, 193}{\$SOURCE_CODE}}). We followed the templates similar to those we found in the artifacts, papers, or technical reports associated with each model. Figure~\ref{fig:vanilla-prompt} shows the three templates used for vanilla prompting of our subject LLMs. Our prompt template for CodeGeeX slightly differs from what is used in their paper~\cite{zheng2023codegeex}. Specifically, their prompt template includes imports, class declaration, and method signature of the translation~\cite{customprompt}. However, this is an \textit{unrealistic approach} because such ground truth does not exist and requires human involvement for each translation. Moreover, the code to be translated (from real-world projects or crafted benchmarks) often contains several methods, making it impossible to use the same template.

We consider a translation successful if it compiles, passes runtime checks, and existing tests pass on the translated code. We do not consider static evaluation metrics such as exact match, syntax match, dataflow match~\cite{ren2020codebleu}, CodeBLEU~\cite{ren2020codebleu}, and CrystalBLEU~\cite{eghbali2022crystalbleu} because our goal is to validate (compile and execute) the translations. Static metrics can also be misleading in code synthesis~\cite{chen2021evaluating}---i.e., LLMs may achieve reasonably high numbers for these metrics, but generate code that cannot be executed due to compilation or runtime errors~\cite{ahmad2021avatar,chen2021evaluating}. 
The last seven columns of Table~\ref{table:dataset} show the detailed results of vanilla prompting of subject LLMs for code translation. We next discuss our key observations.

\begin{description}[leftmargin=*]

    \item[$\bullet$] Except for GPT-4 and StarCoder, all other models performed poorly. The biggest surprise here is CodeGeeX, a model trained explicitly for code translation. We believe this result is because, as mentioned, we excluded 
    information about the translated code (imports, class declaration, and method signature) in the prompt. Such information is typically not available and non-trivial to compute. 
    (To check the correctness of our results, we repeated their experiments with their template and ours, which resulted in the pass@1 dropping from 25.6\% to 0.02\% on their dataset, HumanEval-X.) 

\begin{figure}[t]
\includegraphics[width=\linewidth]{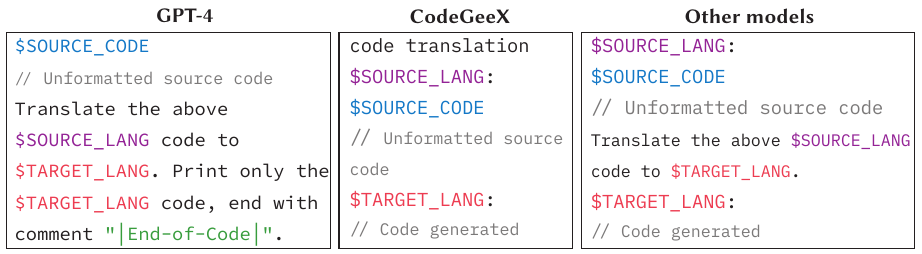}
\vspace*{-18pt}
\caption{Vanilla prompting templates.}
\vspace*{-7pt}
\label{fig:vanilla-prompt}
\end{figure}


    \item[$\bullet$] There is a strong correlation between the average number of tests per translation sample and unsuccessful translation (correlation coefficient $r$ ranging from $0.64$ to $0.85$ for all models). That is, the more rigorous the existing test suite, the better it can evaluate if a translation preserves functionality. 

    \item[$\bullet$] 
    There is no consistent pattern between unsuccessful translations and source/target language, but translating \textit{to} \go results in more compilation errors
    due to its strict syntax constraints, e.g., forbidding unused variables or imports.

    \item[$\bullet$] The LLMs fail to translate real-world projects. This is mainly because crafted benchmark programs are simpler, without complex dependencies or use of language features, e.g., annotations, inheritance, etc. 
    Moreover, in a real-world setting, translating files/methods in isolation, even if successful, may fail at the project level. 
    That said, further manual investigation showed that for the Commons CLI, three out of 22 translated files could be compiled using \texttt{\small py\_compile}~\cite{pycompile}. These simple classes consist of (1) an exception class with only one method, (2) an interface with two method declarations, and (3) a utility class with two simple methods.

\end{description}

\subsection{Outcome of Unsuccessful Translations}
\label{subsec:llmbreakdown}

The previous research question shows that most of the subject LLMs are yet to achieve a reasonable performance for code translation, even on crafted benchmarks, let alone real-world projects. 
At the next step, we were interested to understand if this is due to a lack of understanding of code syntax or semantics by LLM. 
To do this,
we classify unsuccessful translations based on their error outcome: (1) \textit{Compilation Error}, where translated code cannot be compiled, (2) \textit{Runtime Error}, where translated code compiles but fails at runtime with an exception,
(3) \textit{Functional Error}, where the translated code compiles and executes successfully but results in test failure, 
and (4) \textit{Non-terminating Execution}, where the translated code compiles and executes, but does not terminate (encountering an infinite loop or waiting on user input). 

\begin{figure}[t!]
    \centering
    \includegraphics[width=0.85\linewidth]{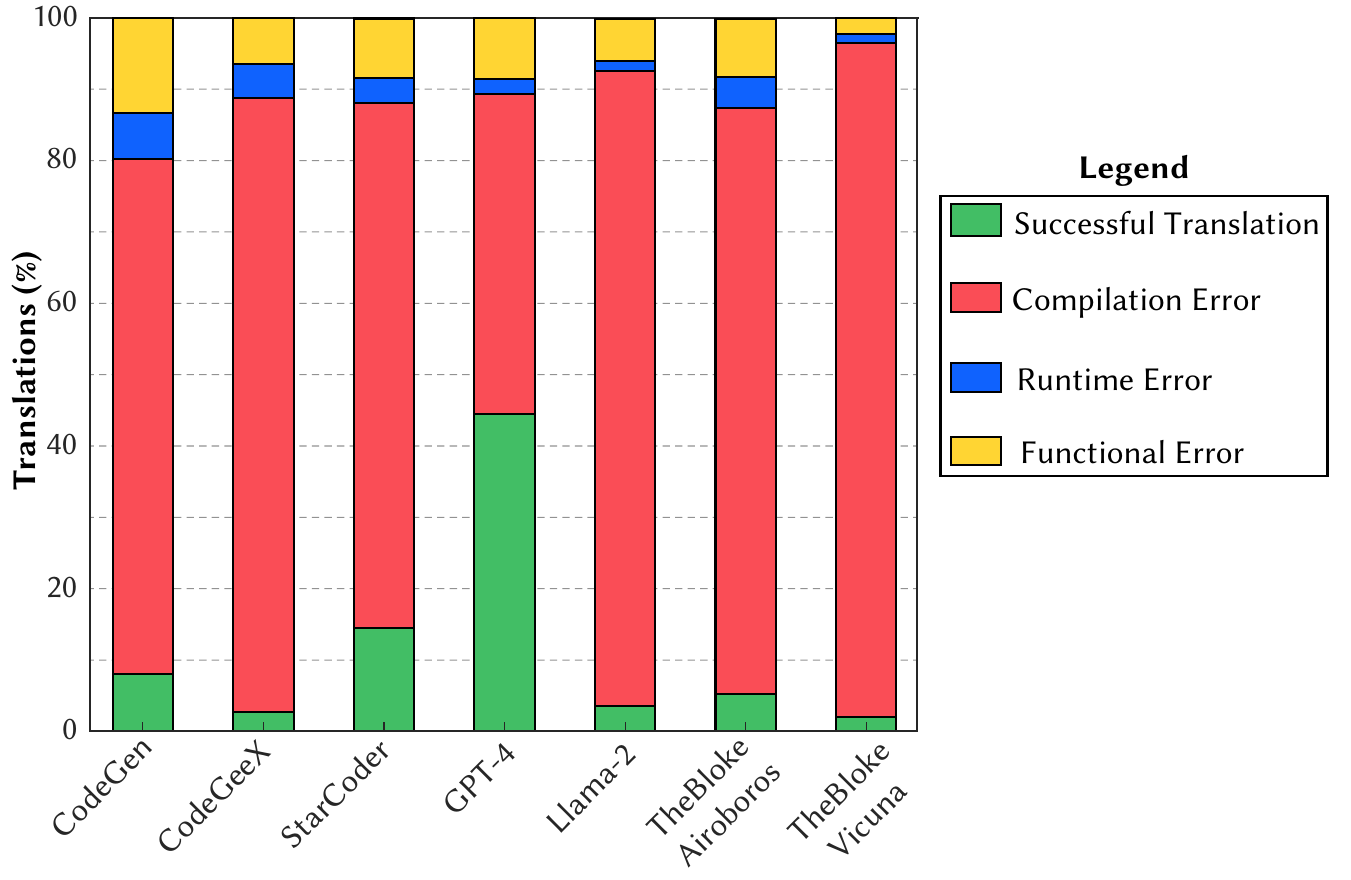}
    \vspace{-1.6em}
    \caption{Outcome of code translations using subject LLMs.}
    \vspace*{-.4em}
    \label{fig:generalization}
\end{figure}

Figure~\ref{fig:generalization} and Table~\ref{table:breakdown} 
show the results of this experiment for each model---accumulated for all subject PLs---and for each subject PL---accumulated for all subject models, respectively. From these results, we observe that most unsuccessful translations result in compilation errors ($77.8\%$), meaning
both general and code LLMs have difficulty understanding code syntax. Further breakdown of the results per PLs shows that \go and \cpp have comparatively stricter syntax, while it is easier for LLMs to generate syntactically correct \py code.\footnote{We used \texttt{py_compile}~\cite{pycompile} to check syntax-related bugs in \py.} The next most common effect of unsuccessful translation is a functional error ($13.4\%$), demonstrating that
often translated code does not preserve the behavior of the source program.




%
\section{LLM-Based Translation Bugs}
\label{sec:taxonomy}
To understand the nature of 
translation bugs, we performed a deep analysis by manually investigating
the root cause of unsuccessful translations. 
Through the following three research questions, we introduce our comprehensive taxonomy of translation bugs (RQ2.1), investigate the prevalence and distribution of each bug category across unsuccessful translations (RQ2.2), and discuss the peculiar characteristics of bugs in real-world projects (RQ2.3). 

\subsection{Taxonomy of Translation Bugs}
\label{subsec:taxonomy}
\begin{figure*}
    \centering
    \includegraphics[width=0.95\linewidth]{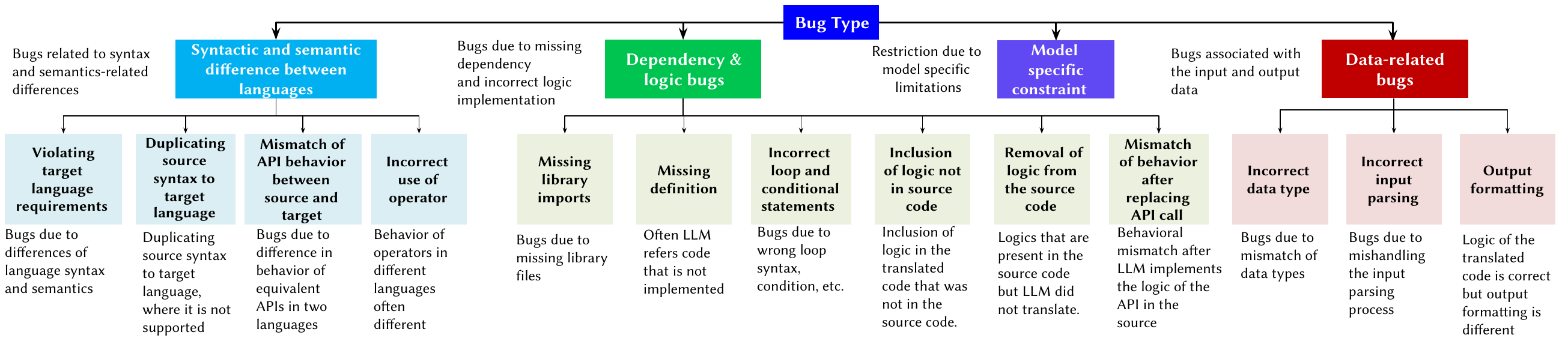}
    \vspace{-1.2em}
    \caption{Taxonomy of bugs introduced while translating code using LLM. \change{The ``other'' category/group is not shown here.} 
    }
    \label{fig:bugtype}
\end{figure*}

Our initial investigation showed that GPT-4 is the best-performing model
and compared to others, its translations exhibit a variety of quality bugs worth investigating. To manage the manual effort
for understanding and labeling bugs for building the taxonomy, we focused on 1,748 unsuccessful translations from GPT-4.

\subsubsection{Methodology}

The manual construction of the taxonomy involved eight human labelers, who are researchers or software engineers in the industry, and involved unsuccessful translations from all 31 translation pairs in Table~\ref{table:dataset}. We built the taxonomy in two phases. In the first phase, we used the \codenet Java-to-Python translations. Each labeler created a taxonomy independently by examining the unsuccessful translations. Then, we combined the individual taxonomies to create a consolidated taxonomy, which served as the initial taxonomy for phase~2. In the second phase, each of the remaining 30 translation pairs was examined by two labelers to assign bug categories to unsuccessful translations. Whenever a new category came up (i.e., a bug could not be covered by the existing categories in the taxonomy), the entire team met to discuss the new category, add it to the taxonomy, and re-label the affected bugs. After completing their labeling tasks, the two labelers assigned to a translation pair met to discuss their labeling, resolve discrepancies, and create the final labeling for the translation pair. The entire exercise took about $630$ person-hours and produced a taxonomy with \change{15}~bug categories organized into \change{five} groups (``model specific constraints'' group does not have any sub-category). In the rest of this section, we discuss the taxonomy groups and bug categories together with illustrative examples.


\subsubsection{Syntactic and semantic differences between the source and the target languages}
\label{sec:syn-sem-diff-bugs}

There are \change{five} bug categories in this group that relate to the failure of an LLM in appropriately handling the syntactic or semantic differences between PLs.

\textbf{A1: Violating target language requirements.} Each PL has its own set of rules that the code must adhere to. For example, in \java, any executable code must be wrapped in a method within a class, whereas in \py, this is not a requirement. Similarly, unused imports in \go result in compilation errors, but not in other PLs.
  
\textbf{A2: Duplicating source syntax to the target language.} LLMs often copy the source PL syntax
even if they are not available in the target PL. In an unsuccessful translation below from \cpp to \go, the subject LLM does not replace \texttt{\small atan2l} API (which is specific to \cpp and does not exist in \go) with an appropriate equivalent. 
        \begin{lstlisting}[language = source,numbers=none]
const ld PI = atan2l(0, -1); # Original C++ code
PI = atan2l(0, -1)           # Incorrect Go code
        \end{lstlisting}
        
\textbf{A3: Mismatch of API behaviors in the source and target.} Library APIs are frequently used in programs in any PL. During translation, API calls in the source need to be either mapped to equivalent API calls in the target PL or implemented from scratch. In the former case, we noted that LLMs often map source APIs incorrectly to the target PL. The following code fragment illustrates such an example where the \java \texttt{\small String.substring()} API is incorrectly mapped to the \go \texttt{\small strings.IndexByte()} API method. 

    \begin{lstlisting}[language = source,numbers=none]
S.substring(i, i + 1)   # Original Java code (returns String) 
strings.IndexByte(S, i) # Incorrect Go code (returns Int64)
    \end{lstlisting}
    
 \textbf{A4: Incorrect use of operators.} The supported operators and their syntax can vary among PLs. For example, \texttt{\small //} in \py represents floor division, for which \java has no corresponding operator. To achieve that behavior in \java, a division must be followed by a call to \texttt{\small Math.floor()}. LLMs can fail in translating such cases.
    \begin{lstlisting}[language = source,numbers=none]
i = i // 10;  # Original Python code (floor division)
i = i / 10;  # Incorrect Java code (division)
    \end{lstlisting}

\vspace{-.5em}
\subsubsection{Dependency and logic bugs in the translated code}
\label{sec:logic-bugs}

These bugs pertain to incorrect dependencies and logic of translated code.

\textbf{B1: Missing library imports.} Import statements are used to load libraries and/or application classes/modules used in code. In several unsuccessful translations, we found that the translated code had 
missing or incorrect imports. 

\textbf{B2: Missing definition.} LLMs can omit definitions/implementation of data types, methods, etc.
In the unsuccessful translation below, the original \cpp code, \texttt{\small main()} calls method \texttt{\small solve()} that is implemented in the same source file. However, after translation to \go, although the call to \texttt{\small solve()} remains, its definition is removed. 
\begin{lstlisting}[language = source,numbers=none]
void solve(){...} # Original C++ code
signed main(){...solve();} # Incorrect Go code
\end{lstlisting}

\textbf{B3: Incorrect loop and conditional statements.} This category covers bugs in translating loops and conditional statements. The following code illustrates the incorrect translation of a \java loop to \py array slicing, resulting in an off-by-one error in the translated code, where the sum excludes the value of \texttt{\small x[200010-k-1]}.
\begin{lstlisting}[language = source,numbers=none, columns=fullflexible]
for(int i = 0; i <= 200010 - k - 1; i++) ans += x[i]; # Java code
ans = sum(x[:200010 - k - 1]) # Incorrect Python code
\end{lstlisting}

\textbf{B4: Inclusion of logic not in source code.} LLMs can generate code that is unrelated to any logic in the source program, thereby, causing the translated code to diverge from the source behavior. In the following example, \texttt{\small max} is initialized to \texttt{\small -1} in the source. However, after translation, it is initialized to a different value; the LLM thus adds logic that does not exist in the source program.
\begin{lstlisting}[language = source,numbers=none, columns=fullflexible]
int max = -1; # Original C code
max_h = max(h) # Incorrect Python code
\end{lstlisting}

\textbf{B5: Removal of logic in the source code.} LLMs often fail to translate part of the source program correctly. For example, in several cases lines such as  \texttt{\#define MAX 101} in \clang are removed.

\textbf{B6: Mismatch of behavior after replacing API call.} These bugs occur when a source API call is translated to custom logic instead of an equivalent API in the target PL (A3 bugs). 
\begin{lstlisting}[language = source,numbers=none, columns=fullflexible]
return Collections.unmodifiableList(requiredOpts); # Java code
return self.required_opts # Incorrect Python code
\end{lstlisting}
The example above from Commons CLI shows an incorrect translation of \java \texttt{\small Collections.unmodifiableList()} that returns an immutable list to \py, where the returned list is mutable.

\subsubsection{Data-related Bugs}
\label{sec:data-bugs}
Data plays an integral role in the code. We found various bugs caused by wrong assumptions about the input data, mismatch of data types, etc.

\textbf{C1: Incorrect data type.} This category of bugs pertains to incorrect types assigned to variables.
The following example, taken from the translation of Commons CLI to \py, illustrates an incorrect type assignment to a field.
\begin{lstlisting}[language = source,numbers=none, columns=fullflexible]
public static Class<File[]> FILES_VALUE=File[].class; # Java code
FILES_VALUE = List[os.path] # Incorrect Python code
\end{lstlisting}

\textbf{C2: Incorrect input parsing.}
Programs that read data from the input stream (i.e., \texttt{\small stdin} or other sources expect the data to be in a specific format. 
Such programs are often translated incorrectly, as illustrated by the following \py-to-\clang translation. The second input line contains three integer values all of which are read into an array in the \py code (line~6), whereas the \clang code reads only two values and assigns~0 for the third value (lines~2--4). This causes the wrong result to be computed in line~6.

 
\noindent\begin{minipage}{.475\linewidth}{
\begin{lstlisting}[language = source,numbersep=5pt]
  -----    Input     -----
2
3 5 2
4 5
------  Source Code ------
N = int(input())
A = [.. input().split()]
...
    d = min(A[i+1], B[i])
\end{lstlisting}}
\end{minipage}\hfill
\begin{minipage}{.475\linewidth}{
\begin{lstlisting}[language = source,numbersep=5pt]
 ----- Translated Code -----
int A[N+1], B[N], N;
for(int i=0; i<N; i++) {
    scanf("%d", &A[i]);}
A[N] = 0;
. . .
    d = A[i+1] < B[i] ?  A[i+1] : B[i];
...
\end{lstlisting}}
\end{minipage}

\textbf{C3: Output formatting.} We found that often, even if the translated code logic is correct, the output is formatted differently. 
In the following example, the source \java code prints `H' followed immediately by an integer value, whereas the translated \py code prints a space between `H' and the integer value. 
\begin{lstlisting}[language = source,numbers=none, columns=fullflexible]
System.out.print("H"); System.out.println(Y - 1988); # Java code
print("H", Y - 1988) # Incorrect Python code
\end{lstlisting}
\vspace{-4pt}
\subsubsection{D: Model-Specific Bugs.} Some bugs are specific to the design of the LLMs used. For instance, having natural language in-between the code, exceeding token size, etc., causing compilation errors or no output to be generated.
We also observed a group of unsuccessful translations---which we refer to as \textit{E: Others}---related to our experiment setup (e.g., memory issues). Given that they do not represent LLM-introduced bugs, we do not include them in the taxonomy.

\begin{table*}[htbp]
  \centering
  \footnotesize
  \setlength\tabcolsep{1.7pt}
  \renewcommand{\arraystretch}{0.61}
  \caption{Types of bugs introduced during code translation by GPT-4 and their occurrences. This table includes all the subject datasets including the real-world projects. All values are in \%.}
   \vspace*{-10pt}
    \begin{tabular}{|l||r|r|r|r||r|r|r|r||r|r|r|r||r|r|r|r||r|r|r|r||r|}
\cline{1-21}    \multicolumn{1}{|c||}{\textbf{Source Language}} & \multicolumn{4}{c||}{\textbf{C}} & \multicolumn{4}{c||}{\textbf{C++}} & \multicolumn{4}{c||}{\textbf{Go}} & \multicolumn{4}{c||}{\textbf{Java}} & \multicolumn{4}{c||}{\textbf{Python}} & \multicolumn{1}{c}{\multirow{2}[3]{*}{\cellcolor[rgb]{ 0,  0,  0}\textcolor[rgb]{ 1,  1,  1}{\textbf{Total}}}} \\
\cline{1-21}    \multicolumn{1}{|c||}{\textbf{Target Language}} & \multicolumn{1}{c|}{\textbf{C++}} & \multicolumn{1}{c|}{\textbf{Go}} & \multicolumn{1}{c|}{\textbf{Java}} & \multicolumn{1}{c||}{\textbf{Py}} & \multicolumn{1}{c|}{\textbf{C}} & \multicolumn{1}{c|}{\textbf{Go}} & \multicolumn{1}{c|}{\textbf{Java}} & \multicolumn{1}{c||}{\textbf{Py}} & \multicolumn{1}{c|}{\textbf{C}} & \multicolumn{1}{c|}{\textbf{C++}} & \multicolumn{1}{c|}{\textbf{Java}} & \multicolumn{1}{c||}{\textbf{Py}} & \multicolumn{1}{c|}{\textbf{C}} & \multicolumn{1}{c|}{\textbf{C++}} & \multicolumn{1}{c|}{\textbf{Go}} & \multicolumn{1}{c||}{\textbf{Py}} & \multicolumn{1}{c|}{\textbf{C}} & \multicolumn{1}{c|}{\textbf{C++}} & \multicolumn{1}{c|}{\textbf{Go}} & \multicolumn{1}{c||}{\textbf{Java}} & \multirow{-2}{*}{\cellcolor[rgb]{ 0,  0,  0}\textcolor[rgb]{ 1,  1,  1}{\textbf{Total}}} \\
\hline
\hline
    \rowcolor[rgb]{ .855,  .933,  .953} A1: Violating target language requirements & 23.1 & 78.2 & 57.7 & 20.0 & 25.9 & 58.5 & 23.5 & 11.3 & 8.3 & 37.5 & 3.6 & 10.0 & 20.4 & 2.7 & 55.8 & 4.6 & 14.5 & 5.8 & 41.4 & 11.2 & \cellcolor[rgb]{ .502,  .502,  .502}\textcolor[rgb]{ 1,  1,  1}{24.3} \\
    \hline
    \rowcolor[rgb]{ .855,  .933,  .953} A2: Duplicating source syntax to the target language & 0.0 & 0.0 & 0.0 & 0.0 & 18.5 & 3.1 & 5.9 & 0.0 & 2.8 & 4.2 & 3.6 & 2.5 & 0.9 & 6.8 & 0.6 & 1.1 & 4.6 & 1.2 & 1.8 & 2.9 & \cellcolor[rgb]{ .502,  .502,  .502}\textcolor[rgb]{ 1,  1,  1}{2.4} \\
    \hline
    \rowcolor[rgb]{ .855,  .933,  .953} A3: Mismatch of API behaviors in the source and target & 0.0 & 1.8 & 0.0 & 0.0 & 0.0 & 0.0 & 0.0 & 0.0 & 11.1 & 0.0 & 0.0 & 2.5 & 4.4 & 5.5 & 3.5 & 16.1 & 0.0 & 2.9 & 0.5 & 1.0 & \cellcolor[rgb]{ .502,  .502,  .502}\textcolor[rgb]{ 1,  1,  1}{3.3} \\
    \hline
    \rowcolor[rgb]{ .855,  .933,  .953} A4: Incorrect use of operator & 0.0 & 0.0 & 0.0 & 2.2 & 0.0 & 0.0 & 2.9 & 0.0 & 0.0 & 0.0 & 3.6 & 0.0 & 0.0 & 1.4 & 0.0 & 0.0 & 1.2 & 0.0 & 0.0 & 2.4 & \cellcolor[rgb]{ .502,  .502,  .502}\textcolor[rgb]{ 1,  1,  1}{0.6} \\
    \hline
    \rowcolor[rgb]{ 0,  .69,  .941} \textcolor[rgb]{ 1,  1,  1}{A: Syntactic and semantic differences between languages} & \textcolor[rgb]{ 1,  1,  1}{23.1} & \textcolor[rgb]{ 1,  1,  1}{80.0} & \textcolor[rgb]{ 1,  1,  1}{57.7} & \textcolor[rgb]{ 1,  1,  1}{22.2} & \textcolor[rgb]{ 1,  1,  1}{44.4} & \textcolor[rgb]{ 1,  1,  1}{61.5} & \textcolor[rgb]{ 1,  1,  1}{32.4} & \textcolor[rgb]{ 1,  1,  1}{11.3} & \textcolor[rgb]{ 1,  1,  1}{22.2} & \textcolor[rgb]{ 1,  1,  1}{41.7} & \textcolor[rgb]{ 1,  1,  1}{10.7} & \textcolor[rgb]{ 1,  1,  1}{15.0} & \textcolor[rgb]{ 1,  1,  1}{25.7} & \textcolor[rgb]{ 1,  1,  1}{16.4} & \textcolor[rgb]{ 1,  1,  1}{59.9} & \textcolor[rgb]{ 1,  1,  1}{21.8} & \textcolor[rgb]{ 1,  1,  1}{20.2} & \textcolor[rgb]{ 1,  1,  1}{9.9} & \textcolor[rgb]{ 1,  1,  1}{43.6} & \textcolor[rgb]{ 1,  1,  1}{17.6} & \cellcolor[rgb]{ .502,  .502,  .502}\textcolor[rgb]{ 1,  1,  1}{30.5} \\
    \hline
    \hline
    \rowcolor[rgb]{ .922,  .945,  .871} B1: Missing library imports & 0.0 & 0.0 & 7.7 & 0.0 & 0.0 & 1.5 & 0.0 & 0.0 & 8.3 & 0.0 & 0.0 & 0.0 & 3.5 & 5.5 & 1.2 & 8.6 & 7.5 & 32.0 & 2.7 & 22.4 & \cellcolor[rgb]{ .502,  .502,  .502}\textcolor[rgb]{ 1,  1,  1}{8.6} \\
    \hline
    \rowcolor[rgb]{ .922,  .945,  .871} B2: Missing definition & 0.0 & 0.0 & 7.7 & 0.0 & 11.1 & 3.1 & 14.7 & 1.9 & 13.9 & 4.2 & 0.0 & 0.0 & 2.7 & 4.1 & 0.0 & 0.0 & 3.5 & 5.8 & 0.5 & 0.0 & \cellcolor[rgb]{ .502,  .502,  .502}\textcolor[rgb]{ 1,  1,  1}{2.4} \\
    \hline
    \rowcolor[rgb]{ .922,  .945,  .871} B3: Incorrect loop and conditional statements & 15.4 & 1.8 & 3.8 & 2.2 & 7.4 & 4.6 & 5.9 & 3.8 & 0.0 & 8.3 & 0.0 & 0.0 & 1.8 & 0.0 & 1.7 & 4.6 & 5.2 & 2.3 & 1.8 & 0.0 & \cellcolor[rgb]{ .502,  .502,  .502}\textcolor[rgb]{ 1,  1,  1}{2.6} \\
    \hline
    \rowcolor[rgb]{ .922,  .945,  .871} B4: Inclusion of logic not in the source & 0.0 & 1.8 & 3.8 & 4.4 & 3.7 & 3.1 & 0.0 & 1.9 & 13.9 & 12.5 & 3.6 & 5.0 & 5.3 & 1.4 & 0.0 & 2.3 & 5.8 & 4.1 & 0.9 & 4.9 & \cellcolor[rgb]{ .502,  .502,  .502}\textcolor[rgb]{ 1,  1,  1}{3.4} \\
    \hline
    \rowcolor[rgb]{ .922,  .945,  .871} B5: Removal of logic from source & 0.0 & 0.0 & 3.8 & 8.9 & 3.7 & 3.1 & 11.8 & 13.2 & 5.6 & 0.0 & 3.6 & 2.5 & 8.8 & 2.7 & 1.2 & 2.9 & 2.3 & 1.7 & 0.9 & 3.4 & \cellcolor[rgb]{ .502,  .502,  .502}\textcolor[rgb]{ 1,  1,  1}{3.3} \\
    \hline
    \rowcolor[rgb]{ .922,  .945,  .871} B6: Mismatch of behavior after replacing API call & 0.0 & 0.0 & 0.0 & 0.0 & 3.7 & 1.5 & 0.0 & 1.9 & 0.0 & 4.2 & 21.4 & 2.5 & 9.7 & 9.6 & 2.9 & 0.6 & 5.2 & 2.3 & 5.0 & 3.9 & \cellcolor[rgb]{ .502,  .502,  .502}\textcolor[rgb]{ 1,  1,  1}{3.8} \\
    \hline
    \rowcolor[rgb]{ 0,  .69,  .314} \textcolor[rgb]{ 1,  1,  1}{B: Dependency \& logic bugs in the translated code} & \textcolor[rgb]{ 1,  1,  1}{15.4} & \textcolor[rgb]{ 1,  1,  1}{3.6} & \textcolor[rgb]{ 1,  1,  1}{26.9} & \textcolor[rgb]{ 1,  1,  1}{15.6} & \textcolor[rgb]{ 1,  1,  1}{29.6} & \textcolor[rgb]{ 1,  1,  1}{16.9} & \textcolor[rgb]{ 1,  1,  1}{32.4} & \textcolor[rgb]{ 1,  1,  1}{22.6} & \textcolor[rgb]{ 1,  1,  1}{41.7} & \textcolor[rgb]{ 1,  1,  1}{29.2} & \textcolor[rgb]{ 1,  1,  1}{28.6} & \textcolor[rgb]{ 1,  1,  1}{10.0} & \textcolor[rgb]{ 1,  1,  1}{31.9} & \textcolor[rgb]{ 1,  1,  1}{23.3} & \textcolor[rgb]{ 1,  1,  1}{7.0} & \textcolor[rgb]{ 1,  1,  1}{19.0} & \textcolor[rgb]{ 1,  1,  1}{29.5} & \textcolor[rgb]{ 1,  1,  1}{48.3} & \textcolor[rgb]{ 1,  1,  1}{11.8} & \textcolor[rgb]{ 1,  1,  1}{34.6} & \cellcolor[rgb]{ .502,  .502,  .502}\textcolor[rgb]{ 1,  1,  1}{24.2} \\
    \hline
    \hline
    \rowcolor[rgb]{ .949,  .863,  .859} C1: Incorrect data type & 7.7 & 1.8 & 0.0 & 11.1 & 7.4 & 15.4 & 26.5 & 1.9 & 27.8 & 4.2 & 7.1 & 7.5 & 8.8 & 8.2 & 7.6 & 0.6 & 21.4 & 15.7 & 8.2 & 21.5 & \cellcolor[rgb]{ .502,  .502,  .502}\textcolor[rgb]{ 1,  1,  1}{11.5} \\
    \hline
    
    \rowcolor[rgb]{ .949,  .863,  .859} C2: Incorrect input parsing & 23.1 & 7.3 & 11.5 & 40.0 & 7.4 & 4.6 & 0.0 & 60.4 & 5.6 & 0.0 & 14.3 & 60.0 & 9.7 & 19.2 & 11.6 & 32.2 & 11.0 & 15.7 & 24.1 & 10.2 & \cellcolor[rgb]{ .502,  .502,  .502}\textcolor[rgb]{ 1,  1,  1}{18.1} \\
    \hline
    \rowcolor[rgb]{ .949,  .863,  .859} C3: Output formatting & 23.1 & 0.0 & 0.0 & 4.4 & 0.0 & 0.0 & 0.0 & 1.9 & 0.0 & 0.0 & 7.1 & 0.0 & 1.8 & 9.6 & 3.5 & 5.2 & 10.4 & 1.7 & 5.0 & 2.4 & \cellcolor[rgb]{ .502,  .502,  .502}\textcolor[rgb]{ 1,  1,  1}{3.9} \\
    \hline
    \rowcolor[rgb]{ .753,  0,  0} \textcolor[rgb]{ 1,  1,  1}{C: Data-related Bugs} & \textcolor[rgb]{ 1,  1,  1}{53.8} & \textcolor[rgb]{ 1,  1,  1}{9.1} & \textcolor[rgb]{ 1,  1,  1}{11.5} & \textcolor[rgb]{ 1,  1,  1}{55.6} & \textcolor[rgb]{ 1,  1,  1}{14.8} & \textcolor[rgb]{ 1,  1,  1}{20.0} & \textcolor[rgb]{ 1,  1,  1}{26.5} & \textcolor[rgb]{ 1,  1,  1}{64.2} & \textcolor[rgb]{ 1,  1,  1}{33.3} & \textcolor[rgb]{ 1,  1,  1}{4.2} & \textcolor[rgb]{ 1,  1,  1}{28.6} & \textcolor[rgb]{ 1,  1,  1}{67.5} & \textcolor[rgb]{ 1,  1,  1}{20.4} & \textcolor[rgb]{ 1,  1,  1}{37.0} & \textcolor[rgb]{ 1,  1,  1}{22.7} & \textcolor[rgb]{ 1,  1,  1}{37.9} & \textcolor[rgb]{ 1,  1,  1}{42.8} & \textcolor[rgb]{ 1,  1,  1}{33.1} & \textcolor[rgb]{ 1,  1,  1}{37.3} & \textcolor[rgb]{ 1,  1,  1}{34.1} & \cellcolor[rgb]{ .502,  .502,  .502}\textcolor[rgb]{ 1,  1,  1}{33.5} \\
    \hline
    \hline
    \rowcolor[rgb]{ .376,  .286,  .478} \textcolor[rgb]{ 1,  1,  1}{D: Model specific constraints} & \textcolor[rgb]{ 1,  1,  1}{7.7} & \textcolor[rgb]{ 1,  1,  1}{7.3} & \textcolor[rgb]{ 1,  1,  1}{3.8} & \textcolor[rgb]{ 1,  1,  1}{4.4} & \textcolor[rgb]{ 1,  1,  1}{11.1} & \textcolor[rgb]{ 1,  1,  1}{1.5} & \textcolor[rgb]{ 1,  1,  1}{8.8} & \textcolor[rgb]{ 1,  1,  1}{1.9} & \textcolor[rgb]{ 1,  1,  1}{0.0} & \textcolor[rgb]{ 1,  1,  1}{0.0} & \textcolor[rgb]{ 1,  1,  1}{25.0} & \textcolor[rgb]{ 1,  1,  1}{0.0} & \textcolor[rgb]{ 1,  1,  1}{8.0} & \textcolor[rgb]{ 1,  1,  1}{4.1} & \textcolor[rgb]{ 1,  1,  1}{9.9} & \textcolor[rgb]{ 1,  1,  1}{14.4} & \textcolor[rgb]{ 1,  1,  1}{2.9} & \textcolor[rgb]{ 1,  1,  1}{2.3} & \textcolor[rgb]{ 1,  1,  1}{5.9} & \textcolor[rgb]{ 1,  1,  1}{7.8} & \cellcolor[rgb]{ .502,  .502,  .502}\textcolor[rgb]{ 1,  1,  1}{6.6} \\
    \hline
    \hline
    \rowcolor[rgb]{ 0,  .125,  .376} \textcolor[rgb]{ 1,  1,  1}{E: Others} & \textcolor[rgb]{ 1,  1,  1}{0.0} & \textcolor[rgb]{ 1,  1,  1}{0.0} & \textcolor[rgb]{ 1,  1,  1}{0.0} & \textcolor[rgb]{ 1,  1,  1}{2.2} & \textcolor[rgb]{ 1,  1,  1}{0.0} & \textcolor[rgb]{ 1,  1,  1}{0.0} & \textcolor[rgb]{ 1,  1,  1}{0.0} & \textcolor[rgb]{ 1,  1,  1}{0.0} & \textcolor[rgb]{ 1,  1,  1}{2.8} & \textcolor[rgb]{ 1,  1,  1}{25.0} & \textcolor[rgb]{ 1,  1,  1}{7.1} & \textcolor[rgb]{ 1,  1,  1}{7.5} & \textcolor[rgb]{ 1,  1,  1}{14.2} & \textcolor[rgb]{ 1,  1,  1}{19.2} & \textcolor[rgb]{ 1,  1,  1}{0.6} & \textcolor[rgb]{ 1,  1,  1}{6.9} & \textcolor[rgb]{ 1,  1,  1}{4.6} & \textcolor[rgb]{ 1,  1,  1}{6.4} & \textcolor[rgb]{ 1,  1,  1}{1.4} & \textcolor[rgb]{ 1,  1,  1}{5.9} & \cellcolor[rgb]{ .502,  .502,  .502}\textcolor[rgb]{ 1,  1,  1}{5.1} \\
    \end{tabular}%
  \label{tb:bugtype}
  \vspace*{-9pt}
\end{table*}%

 \subsection{Prevalence of LLM-based Translation Bugs}
 \label{subsec:prevalent}

We now present the results for RQ2.2, showing the prevalence of bugs in different categories of our taxonomy.
Among the $6197$ attempted translations over 31~language pairs (Table~\ref{table:dataset}), there were $1558$ translation failures. We manually checked and labeled these failures to identify $1748$ bugs. In many cases, an unsuccessful translation has multiple bugs that belong to different categories; in these cases, the translation gets multiple labels.
Table~\ref{tb:bugtype} presents detailed results on the prevalence of translation bugs. In this section, we delve deeper into the characteristics and prevalence of these bugs.
\vspace{-.4em}
\begin{findingBox}{1}{
More than one-third (33.5\%) of the translation bugs are data-related bugs.}
\end{findingBox}
\vspace{-.4em}
As the data for bug group C in Table \ref{tb:bugtype} show, a large proportion of the LLM-introduced bugs is related to data types, parsing input data, and output formatting issues, together accounting for 33.5\% of all bugs. These bugs are particularly prevalent for \py to \clang, \cpp, and \go translations, constituting over 43\%, 33\%, and 37\% of the bugs, respectively, in those translations. Within data-related bugs, our manual investigation found several unique patterns.
\vspace{-.8em}
\begin{findingBox}{1a}
{Among the data-related bugs, most (54\% of the data-related bugs and 18.1\% of the total bugs) are due to incorrect parsing of inputs.}
\end{findingBox}
\vspace{-.8em}
As discussed in \S\ref{sec:data-bugs}, programs that take external inputs contain input-parsing logic, assuming the data to adhere to certain formats, and LLMs often make mistakes while translating this logic; the C2 bug category example in \S\ref{sec:data-bugs} illustrates this.   
A major reason for the prevalence of this bug is that two of our datasets (\codenet and \avatar) consist of programs that read data from the input stream. For \evalplus and the real-world projects, the occurrence of this bug is lesser (one for \evalplus and none for the real-world projects).

\vspace{-.7em}
\begin{findingBox}{1b}
{Choosing the correct data type in the target PL is a crucial step that accounts for 34.3\% of all data-related bugs and 11.5\% of all bugs.}
\end{findingBox}
\vspace{-.5em}
Assignment of correct data types in the translated code (11.5\% of translation bug) is a major challenge. These bugs occur due to incorrect choice of the data type, differences between behaviors of equivalent data types across PLs, and differences in type systems of the source and target PLs.
The example for the C1 bug category in \S\ref{sec:data-bugs} shows an instance of wrong choice of data type in the target PL, where the \java type \texttt{\small Class<File[]>} is converted to list of path objects in \py.
To illustrate an example of equivalent types with different behaviors in source and target PLs, consider the code fragments shown below, where the function \texttt{\small mean_a_d()} in the \py code (left) takes a list of \texttt{\small float} as input. The test case for the function (line~2) uses a large number as test data. The translated code, shown on the right, looks correct and maps \py \texttt{\small float} to the \java \texttt{\small Float} type. However, the equivalent \java test case (line~2 on right) fails because of a fundamental difference between \py \texttt{\small float} and \java \texttt{\small Float}: the former uses 64-bit precision whereas the latter uses 32-bit precision. The \java code thus cannot handle the large test data value, which works fine in \py.

\noindent\begin{minipage}{.47\linewidth}{
\begin{lstlisting}[language = source,numbersep=5pt,showlines=true]
------  Source Code ------
def mean_a_d(numbers: List[float]) -> float: 
self.assertEqual(0.0,mean_a_d([1e+308]))
...
\end{lstlisting}}
\end{minipage}\hfill
\begin{minipage}{.475\linewidth}{
\begin{lstlisting}[language = source,numbersep=5pt]
 ----- Translated Code -----
public static float meanAD(List<Float> n) ...
void testCode() {assertEquals(0.0, meanAD(Arrays.asList(1e+308f)));}
\end{lstlisting}}
\end{minipage}

Incorrect data type bugs can also occur due to differences between the type systems of the source and target PLs. For instance, while translation code in a dynamically typed PL (e.g., \py) to a statically typed PL (e.g., \java), preserving the behavior of source types can be challenging.
\vspace{-.5em}
\begin{findingBox}{2}
~A significant proportion, 30.5\%, of the translation bugs occur due to syntactic and semantic differences between the source and target PLs; almost 80\% of these (24.3\% of all bugs) are caused by violation of target language requirements.
\end{findingBox}
\vspace{-.5em}

Almost one-third of the bugs occur due to reasons, such as violating target language requirements, duplicating source syntax to the target PL, behavioral differences of APIs and operators, etc. \S\ref{sec:syn-sem-diff-bugs} illustrates several bugs in this group. 
For instance, the following translated code violates syntactic constraints of the target language: the variable named \texttt{\small ll} in the \py code (left) is translated verbatim to \cpp, which results in a compilation error as \texttt{\small ll} is a reserved keyword in \cpp (used for long-long data).

\noindent\begin{minipage}{.47\linewidth}{
\begin{lstlisting}[language = source,numbersep=5pt,showlines=true]
------  Source Code ------
ll = - 10 ** 18 - 1
\end{lstlisting}}
\end{minipage}\hfill
\begin{minipage}{.475\linewidth}{
\begin{lstlisting}[language = source,numbersep=5pt]
 ----- Translated Code -----
ll ll = -1e18 - 1;
\end{lstlisting}}
\end{minipage}



Another example of such a bug is the declaration order of methods, where some languages (e.g., \go) permit methods to call another declared subsequently, whereas others (e.g., \cpp) restrict calls to previously declared methods only. Thus, maintaining the wrong declaration order of methods could result in compilation errors.

\vspace{-.5em}
\begin{findingBox}{2a}
{Replacing an API call with another API call in the target PL can result in bugs.}
\end{findingBox}
\vspace{-.5em}
As illustrated for the A3 bug category in \S\ref{sec:syn-sem-diff-bugs}, LLMs can incorrectly map source APIs to the APIs available in the target language.
Table~\ref{tb:bugtype} shows that
%
 3.3\% of all bugs fall in this category.
The following example illustrates an API-mismatch bug, where the LLM replaces the \py \texttt{\small accumulate()} API with \texttt{\small IntStream.concat()} in \java, which can be used in an equivalent manner. However, the LLM erroneously adds \texttt{\small reduce(count).getAsInt()} to reduce the result to an integer value instead of converting the return value of \texttt{\small IntStream.concat()}, an integer stream, to a list/array of integers. This results in an incorrect return value in the two programs: a list of integers in \py and an integer in \java.

\begin{lstlisting}[language = source,numbers=none, columns=fullflexible]
list(accumulate([0] + list(range(1,n)), count)) # Python code
int[] cumsum = IntStream.concat(IntStream.of(0), IntStream.range(1, n)).reduce(count).getAsInt(); # Incorrect Java code
\end{lstlisting}

\vspace{-.5em}
\begin{findingBox}{3}
{24.2\% of the translation bugs are related to incorrect code logic and missing dependencies in the target PL, with missing imports being the dominant category.}
\end{findingBox}
\vspace{-.5em}
LLMs can often translate the source logic incorrectly or miss dependencies. Missing imports is the most frequently occurring bug category (35.5\% of the bugs in the group and 8.6\% of the bugs overall). The other five categories in this group occur in roughly comparable numbers, ranging from 2.4\% to 3.8\% of the overall bugs. 





\subsection{Translation Bugs in Real-world Projects}
 \label{subsec:realworld}

\change{Many of the bugs} seen in the crafted benchmarks translation also occurred while translating Commons CLI and Click. Key among these are: removal of logic in the source (e.g., source methods and field initialization not translated),
inclusion of logic not in source (e.g., implementation added for a stubbed method), missing imports, mismatch in API behaviors
after translation, mismatch of behaviors after replacing API calls 
and incorrect data types. 

However, real-world applications have much more complex code than crafted benchmarks, making them much harder for LLMs to translate. We found nine instances (all from Click) where the translated files contained natural-language text explaining the translation of the source file is infeasible or a non-trivial task for GPT-4. Some others included partial translations of methods in the class, leaving the rest untranslated (translation bug B5). This shows that LLMs, even with longer context windows, cannot capture dependencies between the methods implementing the class logic.
\vspace{-.5em}
\begin{findingBox}{4}
{Real-world applications pose complex challenges for code translation, such as handling method overloading, inheritance, and code annotations, not seen in crafted datasets.}
\end{findingBox}
\vspace{-.5em}
We found examples of language features used in real-world applications that LLMs can struggle with translating. 

\begin{lstlisting}[language = source,numbersep=5pt,showlines=true,columns=fullflexible]
------ Source Code ------
public Options addOption(final Option opt) { ... }
public Options addOption(final String opt, final boolean hasArg, final String description) { ... }
\end{lstlisting}
\vspace{-0.5em}
\begin{lstlisting}[language = source,numbersep=5pt,showlines=true,columns=fullflexible]
----- Translated Code -----
def add_option(self, opt):...
def add_option_arg(self, opt, has_arg, description):...
\end{lstlisting}

For instance, Commons CLI uses method overloading frequently. GPT-4 translates these to Python in different ways, some of which are correct translations, whereas others are erroneous. For an example of the latter, there are cases where GPT-4 translates overloaded Java methods by renaming them in Python to avoid overloaded method names, leaving open the work of suitably renaming all call sites to the methods to preserve the call relations.

In another instance, overloaded methods are translated to methods with the same name, which results in broken functionality because only the last method is available as per the Python semantics, with the previous method definitions overridden.

\noindent\begin{minipage}{.47\linewidth}{
\begin{lstlisting}[language=source,numbersep=5pt,showlines=true,columns=fullflexible]
------ Source Code ------
public static OptionBuilder hasArg() { ... }
public static OptionBuilder hasArg(final boolean hasArg) { ... }
\end{lstlisting}}
\end{minipage}\hfill
\begin{minipage}{.475\linewidth}{
\begin{lstlisting}[language = source,numbersep=5pt,showlines=true,columns=fullflexible]
----- Translated Code -----
@staticmethod
def has_arg():...
@staticmethod
def has_arg(has_arg):
...
\end{lstlisting}}
\end{minipage}

Commons CLI and Click also illustrate the challenges posed by the use of decorators (for adding new functionality to an existing object without modifying its structure) and annotations (for adding metadata to code). For example, Click uses the \texttt{\small @contextmanager} decorator on a method to wrap it with a resource manager. This needs to be translated appropriately in Java to ensure automatic resource release. 
We also observed cases of broken inheritance relations (resulting in missing behaviors/states) in translated code and incorrect translation of exception handling.


\vspace{-.5em}
\begin{findingBox}{5}{
The effectiveness of code translation can vary considerably based on the characteristics of the source and target PLs, such as the type system, available programming APIs, metaprogramming support via decorators or annotations, etc. 
}\end{findingBox}
\vspace{-.5em}

There is a clear pattern of GPT-4 performing much better in translating Commons CLI to Python than Click to Java. This is evident from not only the occurrences of successful translations (three for Commons CLI vs. none for Click) and degenerate code-generation instances (nine for Click vs none for Commons CLI), but also the translation bugs observed. This could be attributed to project-specific complexity characteristics (e.g., the largest source file has $2,436$ NCLOC in Click and $358$ NCLOC in Commons CLI), and it is hard to generalize from limited observations, but language features and the available API ecosystem for a PL can have a considerable impact on the success of code translations. For example, Python-to-Java translations can be more error-prone than Java-to-Python translations, in general, because going from a dynamic type system to a static one can be harder for an LLM to reason about.
\vspace{-.5em}
\begin{findingBox}{6}{
\change{Although the translation bug categories remain the same, occurrences of bug instances and their distribution vary between crafted benchmarks and real-world projects.} 
}\end{findingBox}
\vspace{-.5em}

\change{The bug categories in our taxonomy are broad enough to represent translation bugs in real-world projects. However, their frequencies considerably differ.
The most prevalent bugs in real-world projects are model-specific constraints (29.4\%), missing imports (27.4\%), and the removal of logic from the source code (15.7\%), whereas for crafted benchmarks, data, syntactic, and semantic differences are more common. Moreover, the nature of these bugs is also different. For instance, in most cases, LLM's context window is not large enough to fit both project files and translated code, resulting in model-specific constraint bugs.
Second, LLMs lack specific application knowledge, such as the project structure, external dependencies, and declarations outside the translation scope, i.e., the translated file. The lack of a holistic view of the application can cause bugs such as missing imports and removal of source code logic. Moreover, this limitation can cause inconsistencies in the translated code, e.g., as illustrated for Finding~4, renaming overloaded methods requires all the related call sites to be updated.}

\section{LLM- vs. Non-LLM-based Translation}
\label{sec:comparison}

\begin{table}[tbp]
  \centering
  \footnotesize
  \caption{Successful translation (in \%) of non-LLM and LLM approaches. The top-2 tools are highlighted in teal and violet.}
  \vspace{-8pt}
  \label{table:comparison}
  \setlength\tabcolsep{0.88pt}
    \begin{tabular}{|c|c|c|c|c|c|c|c|c|c|c|c|c|}
    \hline
    \textbf{Dataset} & \textbf{SL}    & \textbf{TL}    & \begin{turn}{65}\textbf{CxGo}\end{turn} & \begin{turn}{65}\textbf{C2Rust}\end{turn} & \begin{turn}{65}\textbf{Java2C\#}\end{turn} & \begin{turn}{65}\textbf{CodeGen}\end{turn} & \begin{turn}{65}\textbf{CodeGeeX}\end{turn} & \begin{turn}{65}\textbf{StarCoder}\end{turn} & \begin{turn}{65}\textbf{GPT-4}\end{turn} & \begin{turn}{65}\textbf{Llama 2}\end{turn} & \begin{turn}{65}\textbf{TB-A}\end{turn} & \begin{turn}{65}\textbf{TB-V}\end{turn} \\
    \hline
    \hline
    \multirow{3}{*}{\codenet} & \multirow{1}{*}{C} &
    Rust &
    -&{\color{teal}\textbf{95.0}}&-
    &
    0.0
    &0.0&10.5&{\color{violet}\textbf{61.0}}&0.5&0.5&0.0
    \\
    \cline{2-13}
    & \multirow{1}{*}{C} & 
    Go &
    {\color{violet}\textbf{62.3}}&-&
    -
    &
    33.0
    &0.0&12.5&{\color{teal}\textbf{72.5}}&5.0&5.5&11.0
    \\
    \cline{2-13}
    & \multirow{1}{*}{Java} & 
    C\# &
    -&-&
    0.0&
    0.5
    &1.0&{\color{violet}\textbf{26.5}}&{\color{teal}\textbf{49.0}}&1.5&1.0&4.0
    \\
    \hline
    \multirow{1}{*}{\avatar} & \multirow{1}{*}{Java} &  C\#
    &-&-&0.0&
    0.0
    &0.0&{\color{violet}\textbf{10.4}}&{\color{teal}\textbf{59.2}}&2.0&0.8&0.4
    \\
    \hline
    \end{tabular}%
  \label{tab:addlabel}\\
  * \{S, T\}L: \{Source, Target\} Language. Data are in \%. TB-\{A, V\}: TB-\{Airoboros, Vicuna\}.
  \vspace{-14pt}
\end{table}%



\sloppy This section compares LLM- and non-LLM-based code translation techniques concerning their effectiveness and the differences in translation bugs. Non-LLM-based techniques are either (1) transpilers (i.e., source-to-source compilers) or cross-lingual code executors for translation (e.g., C2Rust~\cite{c2rust}, CxGo~\cite{c2go}, Sharpen~\cite{sharpen}, and JavaToCSharp~\cite{java2csharp}), or (2) learning-based, which leverage neural machine translation techniques to convert a code from one programming language to another (e.g., mppSMT~\cite{nguyen2015divide} and Tree2Tree~\cite{chen2018tree}). 

Among the mentioned tools, we used CxGo~\cite{c2go} and C2Rust~\cite{c2rust} to translate C code in our \codenet dataset to Go and Rust, and JavaToCSharp~\cite{java2csharp} to translate Java code in \codenet and \avatar datasets to C\#. Accordingly, we used our seven subject LLMs to translate code for all three PL pairs. Table~\ref{table:comparison} illustrates the success rate of 
both LLM and non-LLM approaches.



\textbf{\java to C\#.} We observed that none of the translated code using JavaToCSharp can compile and, in fact, require heavy rewriting of library APIs (e.g., System.in and java.util.Scanner). 
Developers of JavaToCSharp confirmed that this tool is not built to perform full code migration. On the contrary, GPT-4 and StarCoder achieve $54.1\%$ and $18.45\%$ success rate in translating Java code to C\#.  
\vspace{-0.5em}
\begin{findingBox}{7}{
For C to Go, the best-performing LLM, i.e., GPT-4,  achieves 10\% higher success rate than non-LLM-based approach, whereas, for C to Rust, the non-LLM-based approach translates 95\% of the code (34\% better than best-performing LLM). 
}\end{findingBox}
\vspace{-0.5em}

\textbf{C to Go.} CxGo transforms the C code into a common abstract syntax tree (AST) that represents both C and Go constructs. It then converts the AST into Go for translation. When compared to LLMs, we found that CxGo outperforms all the open-source models with an accuracy of 62.25\%, but falls 10\% behind GPT-4. The biggest caveat is that executing code generated by CxGo requires specific libraries, i.e., \texttt{\small stdio} for I/O operations instead of a more generic \texttt{\small fmt} library. In terms of the effect of the incorrect translation, 46.7\% and 53.3\% are due to compilation and runtime errors. Whereas, for the best-performing LLM, i.e., GPT-4, 76.3\%, 12.7\%, and 9.1\% errors are related to compilation, test, and runtime failures.

With a further investigation on the type of translation bugs, we found that most GPT-4 translation bugs (71.8\%) are related to unused imports, variables, and other target language-specific violations, followed by incorrect input parsing (10.3\%). Whereas, with CxGo, the majority of the bugs are related to input parsing (26.7\%) and mismatch of behavior after replacing API call (26.7\%). In most cases, the input parsing related API, i.e., \texttt{scanf}, has been replaced by an API call that uses a custom library written specifically for the tool. Other bugs include missing definitions of API functions that are called from the translated code. The nature of these bugs is significantly different as the non-LLM-based approaches tend to have more bugs related to the custom library, while LLM-based approaches have more syntax and semantics-related bugs.

\textbf{C to Rust.} C2Rust is a transpiler that parses C code into C AST, converts the C AST into equivalent Rust AST by considering the differences in syntax, memory management, and ownership semantics between the two languages, and converts the Rust AST into Rust code. Code translation using C2Rust achieves a significantly higher success rate (95\%) when compared to GPT-4 (61\%).
Using GPT-4, 50\% of the translation bugs were due to compilation errors, 39.3\% were caused by runtime errors, and 10.7\% were due to test failures. As for C2Rust, the 5\% unsuccessful translations were due to compilation errors, the majority of them caused by unused imports. 

\begin{figure}
    \centering
    \vspace{-10pt}
    \includegraphics[width=0.86\linewidth]{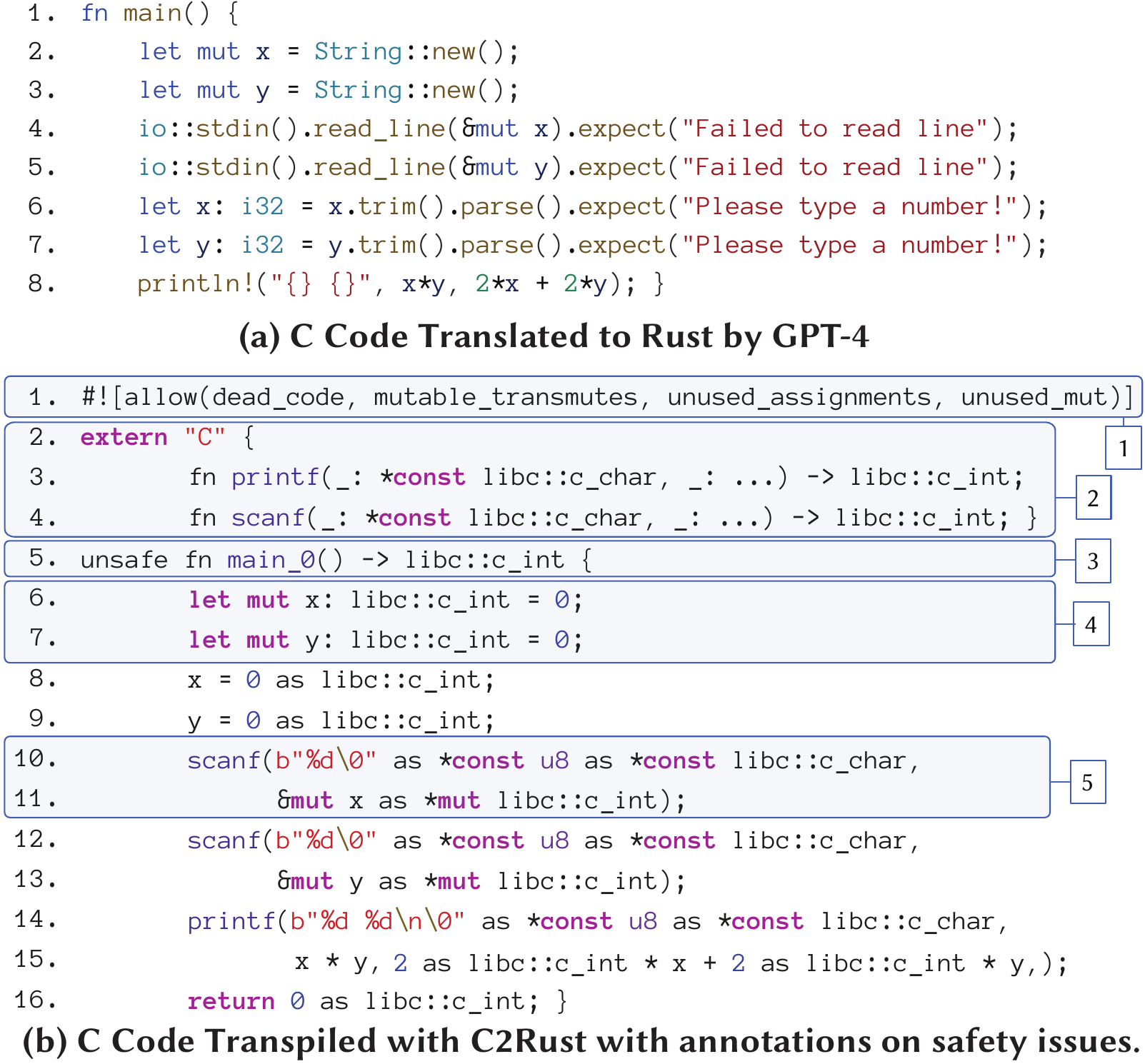}
    \vspace{-1em}
    \caption{Comparative analysis of C-to-Rust code translation.}
   \vspace{-0.5em}
    \label{fig:unsafe-rust}
\end{figure}

\vspace{-0.5em}
\begin{findingBox}{8}{
C2Rust generates non-idiomatic and unsafe code, whereas GPT-4 tends to generate safer and idiomatic code. 
}\end{findingBox}
\vspace{-0.5em}
A close inspection of the translated Rust code offers several noteworthy observations. Figure~\ref{fig:unsafe-rust} shows an example. For the same C code, the translation by GPT-4 (Figure~\ref{fig:unsafe-rust}-(a)) adheres to idiomatic Rust translations, whereas translations produced by C2Rust (Figure~\ref{fig:unsafe-rust}-(b)) contain unsafe code that directs the Rust compiler to bypass any safety checks. In fact, all of the C2Rust generated code are unsafe, whereas, only three GPT-4-translated code are unsafe. 

\begin{itemize}[leftmargin=*, wide=0pt]
    \item \textit{Safety risks from compiler directives}. The use of compiler directives (see \squarebox{1} in Figure~\ref{fig:unsafe-rust}-(b)) can introduce various safety risks: (1) permitting dead code, (2) the \texttt{\small mutable\_transmutes} directive can cause undefined behaviors and mask logical flaws, and (3) allowing unnecessary mutability can cause race conditions. 

    \item \textit{Use of \texttt{\small extern "C"} for foreign function interface}. Potential safety risks may result due to reliance on external C (see \squarebox{2} in Figure~\ref{fig:unsafe-rust}-(b)) functions such as \texttt{\small printf} and \texttt{\small scanf} without Rust's safety guarantees. Specifically, the use of \texttt{\small scanf} (as in \squarebox{5}) without buffer size specifications can lead to vulnerabilities such as buffer overruns.

    \item \textit{Unrestricted use of unsafe block}. The \texttt{\small unsafe} block in \texttt{main} function (see \squarebox{3} in Figure~\ref{fig:unsafe-rust}-(b)) circumvents Rust's safety checks for the code encompassed in that method. This increases the risk of memory safety violations. This issue is further exacerbated by the use of implicit casting (e.g., \texttt{\small 0 as libc::c_int} as in \squarebox{4}) in the method body, which can cause unexpected behavior.
    
\end{itemize}

\vspace{-1em}
\section{Mitigating Translation Bugs}
\label{sec:prompt-engineering}

In this section, we discuss how context information pertaining to unsuccessful translations can help fix buggy translations.

\begin{figure}
    \centering
    \includegraphics[width=0.74\linewidth]{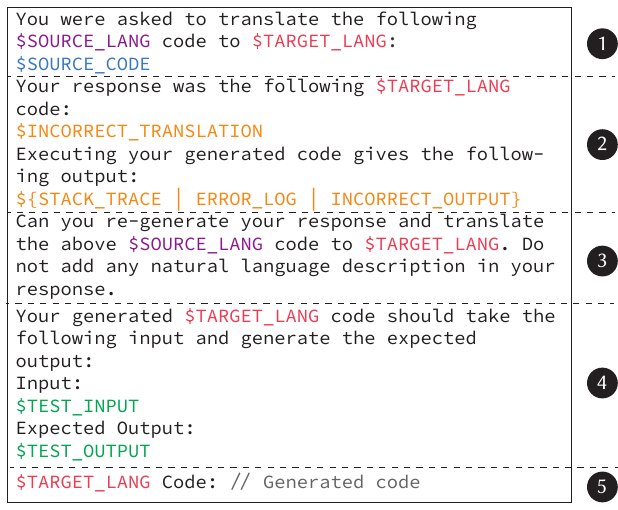}
    \vspace{-10pt}
    \caption{
    Prompt crafting template for LLMs with the context corresponding to the outcomes of unsuccessful translation. 
    }
    \vspace{-1em}
    \label{fig:prompt-template}
\end{figure}

\textit{\textbf{Prompt Crafting}. }
\label{sec:effect-specific-prompts}
Inspired by other works on prompt crafting for fixing bugs~\cite{vaithilingam2022expectation, xia2022less} and how human developers would address a translation error, we propose an \textit{iterative} prompting approach. \change{Our hypothesis is that providing more context information to LLMs can help generate better code.} 
Based on this hypothesis, we include the following contextual information in the revised prompts (Figure~\ref{fig:prompt-template}).

\noindent \circled{1}~\textbf{Source code and original prompt.} Here, we include the original code and the previous prompt used for translating the original code to remind LLMs about the previous task.

\noindent \circled{2}~\textbf{Incorrect translation and error details.} Here, we provide the incorrectly translated code (\texttt{\small\textcolor[RGB]{239,154,69}{\$INCORRECT_TRANSLATION}}), and details regarding the outcome of the translation. If it is runtime error, we provide stack trace (\texttt{\small\textcolor[RGB]{239,154,69}{\$STACK_TRACE}}); for compilation error, we provide error log (\texttt{\small\textcolor[RGB]{255,48,39}{\$ERROR_LOG}}); for test failures, we provide the incorrect output (\texttt{\small\textcolor[RGB]{255,48,39}{\$INCORRECT_OUTPUT}}); finally, for non-terminating execution, we provide a custom message ``The program enters infinite loop.''


\noindent \circled{3}~\textbf{Instructions for translation.} Here, we ask LLM to mitigate the bug and avoid including 
natural-language text in the response.

\noindent \circled{4}~\textbf{Expected behavior.} This part is used optionally if the prior translation was a functional error. Here, we provide test input and expected output pair for the previously wrong output. 
    
\noindent \circled{5} \textbf{Model-specific keyword.} This part is specific to code LLMs and contains the name of the target PL following code LLM templates. 

\noindent All the prompts that we used are in the replication package~\cite{website}.
\begin{figure}
    \centering
    \vspace{-2pt}
    \includegraphics[width=0.7\linewidth]{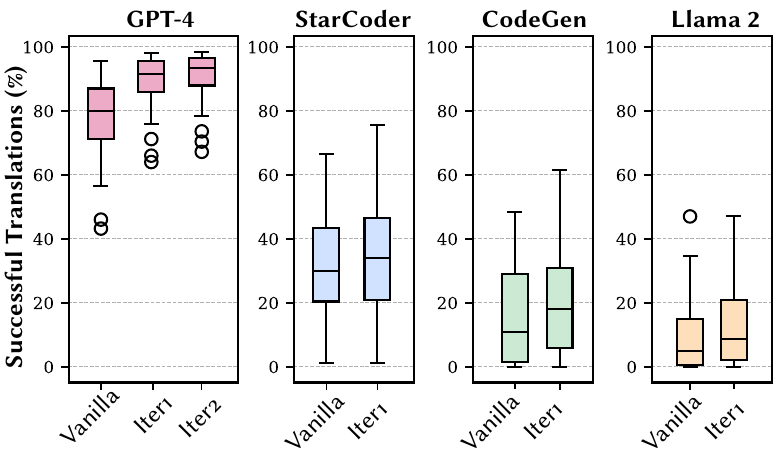}
    \vspace{-1.1em}
    \caption{Effectiveness of providing error-related information in the prompt for fixing bugs.} 
    \vspace{-2pt}
    \label{fig:fixing}
\end{figure}

\textit{\textbf{Iterative Translation Bug Mitigation.}} Our prompt-crafting technique is \textit{iterative}. 
At each iteration $iter_i$, we update the prompt template (Figure~\ref{fig:prompt-template}) with information corresponding to the previously failed translation. We refer to the outcome of $iter_i$ translation as \textit{ 
translation patch}.
At the end of each iteration, we verify if the patch results in a successful translation. If not, we utilize the outcome of the patch and build the prompt for the next iteration.
The iterative mitigation can continue for a fixed number of iterations or until the percentage of successful translations at the previous iteration is smaller than a pre-defined threshold.

To evaluate the effectiveness of our iterative prompt crafting,
we attempted to mitigate unsuccessful translations from RQ1 for four subject LLMs: CodeGen, StarCoder, GPT-4, and Llama 2. We excluded CodeGeeX because its prompt template is rigid, and we could not introduce additional contexts. Due to inferior performance in vanilla prompting, we also excluded the TB-Airoboros and TB-Vicuna. We set the termination criteria so that the mitigation process terminates if the overall increase of successful translation is less than 5\%.
Figure~\ref{fig:fixing} summarizes our findings. 
Based on the proposed termination criterion, our mitigation process lasted for two iterations for GPT-4, and one iteration for the rest of the models.  
The results suggest that the proposed technique can increase the number of successful translations for all the studied LLMs, with $iter_1$---prompting on results from vanilla prompts---increasing the success rate of GPT-4, StarCoder, Codegen, and Llama 2 by 12.33\%, 3.55\%, 2.65\%, 1.97\%, respectively. $iter_2$---prompting on results from $iter_1$, can improve GPT-4 by 1.7\%.
\begin{figure}[tp!]
    \centering
    \includegraphics[width=0.8\linewidth]{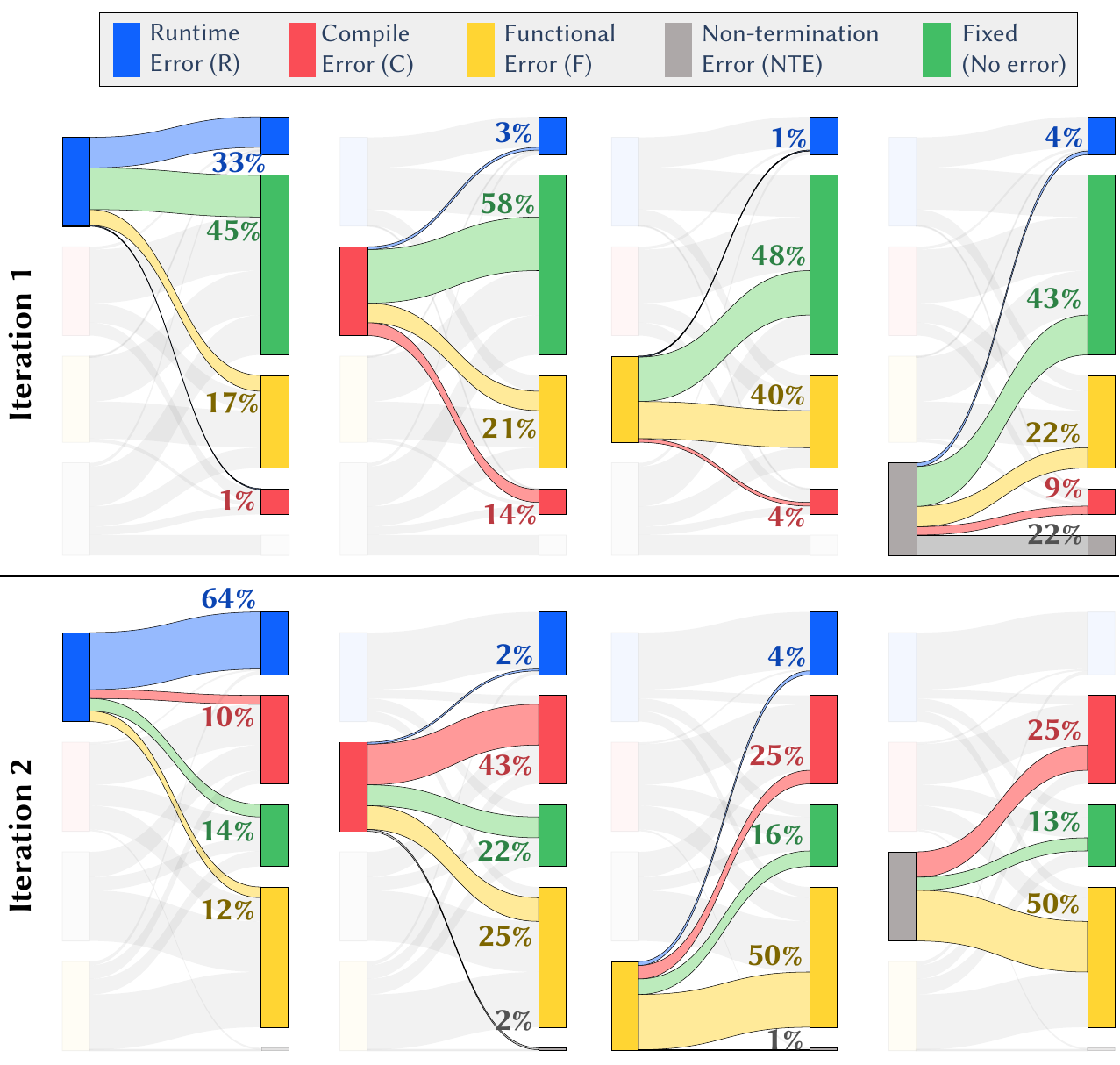}
    \vspace*{-1.1em}
    \caption{Translation outcomes after prompting GPT-4.}
     \vspace*{-.9em}
\label{fig:promptengg}
\end{figure}
We also wanted to understand how translation bugs evolve during this iterative prompting process. To that end, we tracked the error outcomes of unsuccessful translations (\S\ref{subsec:llmbreakdown}) from vanilla prompting to $iter_1$ (for all models) and from $iter_1$ to $iter_2$ (for GPT-4). Figure~\ref{fig:promptengg} shows our analysis results on GPT-4 (other models' results are available in the artifact~\cite{website}).
 
With $iter_1$, we observe a substantial reduction in \textit{compilation errors} for GPT-4, with $58\%$ completely fixed (no error) and $38\%$ transformed to other translation bugs. However, for other errors, the percentage is lower---45\% runtime errors, 48\% functional errors, and 43\% non-terminating executions---suggesting these bugs are harder to mitigate. With $iter_2$, the outcome of most bugs does not evolve and they remain the same. In both cases, we observe a few cases where the outcome of the translation degrades: i.e., functional error transforms to runtime/compilation error, or runtime error transforms to compilation error.
Also, we found several instances where, instead of fixing the bug, the LLM introduced a new one. 
These findings show that, although the proposed technique improves overall effectiveness, future work should be directed toward \change{combining the power of program analysis and more nuanced LLM approaches, e.g., by including more suitable code examples for in-context learning.}
\vspace{-1pt}
\section{Discussion}
\label{sec:discussion}
\begin{table}[t]
  \centering
  \footnotesize
  \caption{Characteristics of LLM-based and non-LLM-based code translation approaches.}
  \vspace{-10pt}
   \setlength\tabcolsep{1.2pt}
    \begin{tabular}{|l|c|c|c|}
    \hline
    \multicolumn{1}{|c|}{\textbf{Characteristics}} & \textbf{Non-LLM} & \textbf{LLM} \\
    \hline

Broader context & \textcolor{teal}{\textbf{\checkmark}}      & \textcolor{red}{\textbf{X}}  \\
Determinism and reasoning & \textcolor{teal}{\textbf{\checkmark}}      & \textcolor{red}{\textbf{X}}  \\
Leveraging target-language idioms & \textcolor{red}{\textbf{X}}     & \textcolor{teal}{\textbf{\checkmark}} \\
    \hline
    \end{tabular}%
  \label{tb:proscons}\\
  \vspace{-16pt}
\end{table}%

\textit{\textbf{Pros and cons of LLM- and non-LLM-based translation.}}
We discuss strengths and weaknesses of each approach with respect to three key characteristics (Table~\ref{tb:proscons}). First, non-LLM-based approaches, specifically transpilers, have a more comprehensive context of an application. On the other hand, while translating a module, the variable types, method signatures, folder structure, and dependencies associated with that module are often unavailable to LLMs because of their restriction on the token size. 
Second, non-LLM-based techniques are generally deterministic, allowing for more predictable reasoning, whereas LLMs are inherently probabilistic, with a greater degree of creativity.
The third characteristic pertains to the naturalness of the translated code. In this respect, non-LLM-based approaches mostly do not leverage target language idioms, which can negatively affect code readability and maintainability. In some cases, this even leads to security issues (e.g., C to Rust). Conversely, LLMs tend to generate more natural, human-like code. More research on utilizing both approaches would be fruitful. 

\noindent\textit{\textbf{Translating real-world projects.}}
Unlike crafted benchmarks, files in real-world projects do not exist as standalone programs, and providing relevant context about inter-file dependencies can help an LLM produce better code. More fundamentally, however, new techniques are required to enable code translation to scale to real-world applications and also generate high-quality translations.
For instance, such techniques could use program analysis to provide more context while translating a piece of code.
Another potential direction is to leverage program-decomposition techniques to split the source file into smaller fragments, each translated separately via prompts that encode appropriate context information about the fragment's dependencies; the translated fragments are then composed to produce the fully translated code.
Finally, a significant challenge in translating real-world projects is handling library API calls and the differences in the API ecosystem of different PLs. There can be cases where no suitable API exists in the target PL to translate an API call to---techniques that combine code summarization and code synthesis could be investigated to fill such gaps.

\noindent\textit{\textbf{{Improving open and closed-source LLMs.}}}~%
The findings of this work shows considerable scope for improving open-source and close-source LLMs for code translation. 
For closed-source models (i.e., GPT-4), the increasing complexity of use cases (i.e., translating real-world projects) calls for an evolution in prompting strategies \change{perhaps with the help of program analysis}.
One promising research direction could involve creating prompts that build upon each other 
\cite{wei2022chain, yao2023tree}, 
following a coherent and logical flow of information.
For open-source LLMs, enhancing the performance could involve fine-tuning~\cite{howard2018universal, sun2019fine}, wherein different models~\cite{jacobs1991adaptive} (instead of one model fits all) are carefully trained to tackle distinct aspects of the translation process (particularly those related to the bug categories). 

\vspace{-5pt}
\section{Threats to Validity}
\label{sec:threat}

Like any empirical study, there are threats to the validity of our results. We discuss the most significant ones of these, along with the mitigating factors.

\noindent\textbf{\textit{External Validity.}} Threats to external validity pertain to whether our results can generalize to other experimental settings. Key factors here include the PLs, LLMs, and datasets selected. To mitigate these threats, we selected five PLs, guided by PL popularity~\cite{tiobe} while also covering different programming paradigms. In terms of datasets, we used multiple well-known datasets with different characteristics and also two real-world projects. For LLM selection, we included several SOTA general and code LLMs. Finally, for non-LLM-based approaches, we used multiple tools covering different PLs.


\noindent\textbf{\textit{Internal Validity.}} As for threats to internal validity, one potential threat is that, for each translation task, we performed the translation once.
Performing a translation task multiple times may change the success rates in translation as LLMs are inherently non-deterministic. However, this does not affect the primary goal of our work, which is to study the characteristics of translation bugs. 
Regardless of randomness or number of repetitions, the nature of an unsuccessful translation remains unchanged.
Moreover, to reduce the effects of LLMs' sensitivity to prompt templates, we followed the best practices described in the respective artifacts/papers/reports (refer \S\ref{sec:empirical-setup}).
Another threat to internal validity is that we do not have a formal inter-rater reliability metric for our manual labeling of bugs. To mitigate this threat, after each labeling round, both labelers met to discuss any discrepancies and resolve them. At the end, each bug was assigned a single mutually agreed label. On top of that, one author went through the entire labeling to ensure consistency among the groups. Finally, our results may be affected by bugs in our automation scripts. To address this threat, we thoroughly tested the scripts and spot-checked the results for correctness. Moreover, we make our artifacts publicly available~\cite{website} to enable review and replication of our results.

\noindent\textbf{\textit{Construct Validity.}} We measured translation correctness using the test cases provided in our datasets, which is a commonly used approach for assessing the quality of generated code. This approach comes with the risk that inadequate or weak test suites can cause buggy translations that pass the test suites to be considered correct. We note that one of our datasets, CodeNet~\cite{puri2codenet}, contains one test case per sample and may be susceptible to this threat. However, the other datasets used contain fairly rigorous test suites. 

%
\vspace{-5pt}
\section{Related Work}
\label{sec:related-work}

\textbf{\textit{Code Translation and Synthesis.}}
\change{There exist two broad classes of code-translation approaches. The first category includes tools such as transpilers, or source-to-source compilers, that leverage program-analysis techniques for converting code from one PL to another.
For instance, C2Rust~\cite{c2rust} and CxGo~\cite{c2go} are transpilers that translate C programs to Rust and Go, respectively, whereas Sharpen~\cite{sharpen} and Java2CSharp~\cite{j2csharp} convert Java code to C\#. The second category includes learning-based techniques, including lexical statistical machine translation~\cite{nguyen2013lexical, nguyen2014migrating} and tree-based neural networks~\cite{chen2018tree} for translating Java to C\#. Other work in the category leverages deep learning and unsupervised learning~\cite{roziere2020unsupervised, lachaux2021dobf} for translating C++, Java, and Python code, phase-based statistical learning~\cite{karaivanov2014phrase} for C\#-to-Java translation, etc.} More recent techniques leverage LLMs (e.g., StarCoder~\cite{starcoder}, PolyCoder~\cite{polycoder}, SantaCoder~\cite{santacoder}, CodeGen~\cite{nijkamp2022codegen}, BLOOM~\cite{bloom}, CodeT5~\cite{codet5}, CodeX~\cite{chen2021evaluating}, GPT-4~\cite{gpt4tr}, Llama 2~\cite{llama2}, etc.) for code generation and code translation (CodeGeeX~\cite{zheng2023codegeex}). 
However, in the context of LLMs, there is no study that understands the bugs introduced by LLMs during code-translation tasks. 

\noindent\textbf{\textit{Bug and repair study.}} Software bugs are well studied~\cite{beizer1990software} including several works on deep learning model-related bugs~\cite{islam2019comprehensive, zhang2018empirical, islam2020repairing}. 
Moreover, there is also an extensive list of works on fixing software bugs using LLMs (e.g., ~\cite{xia2022less, feng2020codebert, xia2023conversational, joshi2023repair},
and other approaches (e.g.,~\cite{xia2022practical,zhang2023survey,gazzola2019automatic}). 
\change{Compared to both of these classes of studies, here, we study bugs introduced by LLMs while translating code from one PL to another. Also, we compare the effectiveness of LLM with non-LLM-based approaches in terms of code translation task.}

%
\vspace{-5pt}
\section{Concluding Remarks}
\label{sec:conclusion}

Code translation has various applications, from modernizing enterprise applications to migrating legacy software to modern PLs. Given the promising performance of LLMs in code synthesis, 
we were interested to understand how they perform in code translation task.
Our empirical investigation of the general and code LLMs across five PLs and several benchmarks and real-world projects demonstrates that state-of-the-art LLMs are yet to effectively automate code translation, specifically to translate complex real-world projects. Through meticulous manual analysis, we also identified $15$ root causes of unsuccessful translations. 
\change{Furthermore, to investigate how providing more context can help LLMs generate better code, we presented and evaluated an iterative prompt-crafting technique. We also assessed existing non-LLM-based techniques, comparing their strengths and weaknesses.}
We are currently considering several directions for future work. First, we want to investigate \change{how existing SE/PL techniques can help mitigate these bugs, including giving more context to LLMs}. Second, our ultimate goal is to advance real-world application translation. 

\begin{acks}
This work is supported by IBM-Illinois Discovery Accelerator Institute and NSF CCF 22-38045 CAR grants. We thank Darko Marinov, David Grove, John Rofrano, Maja Vuković, and Seetharami Seelam for their help with this research. We also thank the anonymous reviewers for their comments, which helped make this work stronger.
\end{acks}

\balance
\bibliographystyle{ACM-Reference-Format}
\bibliography{reference}

\end{document}